\documentclass[12pt, draftclsnofoot, onecolumn]{IEEEtran}
\IEEEoverridecommandlockouts 
\ifCLASSINFOpdf
\usepackage[pdftex]{graphicx}
\graphicspath{{./figure/ }}
\DeclareGraphicsExtensions{.pdf,.jpeg,.png}
\else
\usepackage[dvips]{graphicx}
\DeclareGraphicsExtensions{.eps}
\fi

\usepackage{tikz}
\usepackage{pgfplots}
\pgfplotsset{compat=1.15}
\usetikzlibrary{calc,shapes.geometric,shapes.misc,arrows,matrix}
\usepackage[absolute,overlay]{textpos}

\usepackage[font=footnotesize]{caption}
\usepackage[font=footnotesize]{subcaption}
\usepackage{float}
\graphicspath{{figure/}}

\usepackage[utf8]{inputenc}
\usepackage[nolist]{acronym}
\usepackage[cmex10]{amsmath}
\usepackage{mdwmath}
\usepackage{mdwtab}
\usepackage{afterpage}
\usepackage{multirow}
\usepackage{amsthm}
\usepackage{amssymb}
\usepackage{mathtools}

\newtheorem{proposition}{Proposition}
\newtheorem{lemma}{Lemma}

\usepackage{siunitx}
\DeclareSIUnit{\dBi}{dBi}
\DeclareSIUnit{\dBm}{dBm}
\DeclareSIUnit{\dBW}{dBW}
\usepackage{color}
\usepackage{dashrule}
\usepackage{algorithm} 
\usepackage{algpseudocode}
\usepackage{cite}
\floatstyle{plaintop}
\restylefloat{table}
\usepackage{balance}
\usepackage{bm}
\usepackage{xspace}

\usepackage{amssymb}

\newcommand{\vek}[1]{\ensuremath{\mathbf{#1}}}          
\newcommand{\vekt}[1]{\ensuremath{\tilde{\vek{#1}}}}        

\newcommand{\lagfunc}{\ensuremath{\mathcal{L}}}
\newcommand{\lagrange}{\ensuremath{\omega}}          

\newcommand{\E}{\ensuremath{\mathbb{E}}}

\newcommand{\nvar}{\ensuremath{\sigma_\mathsf{n}^2}}
\newcommand{\naltvar}{\bar{\sigma}_{\mathsf{n}}^2}
\newcommand{\naltvarl}{\bar{\sigma}_{\mathsf{n},\ell}^2}

\newcommand{\NS}{\ensuremath{N_{\mathrm{S}}}}          

\newcommand{\Nt}{\ensuremath{N_{\mathrm{t}}}}
\newcommand{\Nr}{\ensuremath{N_{\mathrm{r}}}}
\newcommand{\NTx}{\ensuremath{N_{\mathrm{Tx}}}}

\newcommand{\GTx}{\ensuremath{\zeta_{\mathrm{Tx,dB}}}}
\newcommand{\GRx}{\ensuremath{\zeta_{\mathrm{Rx,dB}}}}

\newcommand{\DATx}{\ensuremath{D_{\mathrm{Tx}}}}
\newcommand{\DARx}{\ensuremath{D_{\mathrm{Rx}}}}
\newcommand{\DS}{\ensuremath{D_{\mathrm{S}}}}

\newcommand{\Ptx}{\ensuremath{P_{\text{Tx}}}}

\newcommand{\aoa}{\ensuremath{\theta}}
\newcommand{\aoal}{\ensuremath{\theta_\ell}}
\newcommand{\aod}{\ensuremath{\Theta}}
\newcommand{\aodl}{\ensuremath{\Theta_\ell}}

\newcommand{\erraoa}{\ensuremath{\upsilon}}

\newcommand{\estaod}{\ensuremath{\hat{\aod}_{\ell}}}
\newcommand{\erraod}{\ensuremath{\xi}}

\newcommand{\steeraod}{\ensuremath{\vek{b}}}
\newcommand{\steeraodest}{\ensuremath{\hat{\steeraod}}}
\newcommand{\Steeraod}{\ensuremath{\vek{B}}}
\newcommand{\steeraoa}{\ensuremath{\vek{a}}}
\newcommand{\steeraoaest}{\ensuremath{\hat{\steeraoa}}}
\newcommand{\Steeraoa}{\ensuremath{\vek{A}}}

\newcommand{\pdfaoa}{\ensuremath{f_{\mathsf{\upsilon}}}}

\newcommand{\cfaoa}{\ensuremath{\varphi_{\mathsf{\upsilon}}}}
\newcommand{\cfaod}{\ensuremath{\varphi_{\mathsf{\xi}}}}
\newcommand{\cfgau}{\ensuremath{\varphi_{\mathsf{G}}}}
\newcommand{\cfuni}{\ensuremath{\varphi_{\mathsf{U}}}}

\newcommand{\corrmtx}{\ensuremath{\vek{R}}}

\newcommand{\RE}{\ensuremath{r_{\text{E}}}}

\newcommand{\C}{\ensuremath{\mathbb{C}}}

\newcommand{\D}{\ensuremath{\text{D}}}

\newcommand{\fc}{\ensuremath{f_{\text{c}}}}

\newcommand{\veka}{\ensuremath{\vek{a}}}
\newcommand{\vekb}{\ensuremath{\vek{b}}}
\newcommand{\vekx}{\ensuremath{\vek{x}}}
\newcommand{\veks}{\ensuremath{\vek{s}}}

\newcommand{\veky}{\ensuremath{\vek{y}}}

\newcommand{\vekn}{\ensuremath{\vek{n}}}

\newcommand{\EQ}{\ensuremath{\vek{W}}}
\newcommand{\eq}{\ensuremath{\vek{w}}}
\newcommand{\eqsat}{\ensuremath{\vek{w}_{\ell}}}
\newcommand{\PC}{\ensuremath{\vek{G}}}
\newcommand{\pc}{\ensuremath{\vek{g}}}

\newcommand{\vekH}{\ensuremath{\vek{H}}}

\newcommand{\vekHsat}{\ensuremath{\vek{H}_{\ell}}}

\newcommand{\chtx}[1]{\ensuremath{\vekH_{#1}}}

\newcommand{\chest}{\ensuremath{\vekt{H}}}

\newcommand{\vekI}{\ensuremath{\vek{I}}}

\newcommand{\vekS}{\ensuremath{\vek{S}}}

\newcommand{\vekR}{\ensuremath{\vek{R}}}

\newcommand{\vekP}{\ensuremath{\vek{P}}}
\newcommand{\vekZ}{\ensuremath{\vek{Z}}}

\newcommand{\vekV}{\ensuremath{\vek{V}}}
\newcommand{\vekSig}{\ensuremath{\boldsymbol{\Sigma}}}
\newcommand{\vektU}{\ensuremath{\vekt{U}}}
\newcommand{\vektV}{\ensuremath{\vekt{V}}}
\newcommand{\vektSig}{\ensuremath{\boldsymbol{\tilde{\Sigma}}}}
\newcommand{\vekA}{\ensuremath{\vek{A}}}

\newcommand{\dB}{\ensuremath{\text{dB}}}

\newcommand{\diag}[1]{\ensuremath{\operatorname{diag}\left(#1\right)}}
\newcommand{\blkdiag}[1]{\ensuremath{\operatorname{blkdiag}\left(#1\right)}}
\newcommand{\tr}[1]{\ensuremath{\operatorname{tr}\left\{#1\right\}}} 
\newcommand{\rank}[1]{\ensuremath{\operatorname{rank}\left(#1\right)}}
\newcommand{\sinc}[1]{\ensuremath{\operatorname{sinc}\left(#1\right)}}

\newcommand{\nullvec}{\boldsymbol{0}}   

\begin{acronym}
	\acro{rx}[GS]{ground station}
	\acro{dl}[DL]{downlink}
	\acro{ul}[UL]{uplink}
	\acro{mimo}[MIMO]{multiple-input-multiple-output}
	\acro{ntn}[NTN]{non-terrestrial network}
	\acro{geo}[GEO]{geostationary orbit}
	\acro{meo}[MEO]{medium Earth orbit}
	\acro{leo}[LEO]{low Earth orbit}
	\acro{qos}[QoS]{quality of service}
	\acro{dip}[DiP]{distributed precoding}
	\acro{pspc}[PAPPC]{per-acces point power constraint}
	\acro{kkt}[KKT]{Karush-Kuhn-Tucker}
	\acro{svd}[SVD]{singular value decomposition}
	\acro{mse}[MSE]{mean-squared error}
	\acro{mmse}[MMSE]{minimum mean-squared error}
	\acro{mvdr}[MVDR]{minimum variance distortionless response}
	\acro{zf}[ZF]{zero-forcing}
	\acro{mrt}[MRT]{maximum ratio transmission}
	\acro{mrc}[MRC]{maximum ratio combining}
	\acro{sic}[SIC]{successive interference cancellation}
	\acro{iid}[i.\,i.\,d.]{independent and identically distributed}
	\acro{snr}[SNR]{signal-to-noise ratio}
	\acro{sinr}[SINR]{signal-to-interference-and-noise ratio}
	\acro{csi}[CSI]{channel state information}
	\acro{csit}[CSIT]{\ac{csi} at the transmitter}
	\acro{csir}[CSIR]{\ac{csi} at the receiver}
	\acro{los}[LOS]{line-of-sight}
	\acro{nlos}[NLOS]{non-\ac{los}}
	\acro{pl}[PL]{path loss}
	\acro{fspl}[FSPL]{free space \ac{pl}}
	\acro{aod}[AoD]{angle of departure}
	\acrodefplural{aod}{angles of departure}
	\acro{aoa}[AoA]{angle of arrival}
	\acrodefplural{aoa}{angles of arrival}
	\acro{ula}[ULA]{uniform linear array}
	\acro{ura}[URA]{uniform rectangular array}
	\acro{eci}[ECI]{Earth-centered inertial}
	\acro{ecef}[ECEF]{Earth-centered, Earth-fixed }
	\acro{dft}[DFT]{discrete Fourier transform}
	\acro{sota}[SotA]{state-of-the-art}
	\acro{pdf}[PDF]{probability density function}
	\acro{cf}[CF]{characteristic function}
	\acro{fdd}[FDD]{frequency division duplex}
\end{acronym}

\allowdisplaybreaks[4]

\DeclareMathOperator*{\argmax}{arg\,max}

\allowdisplaybreaks

\usepackage{moreverb}
\immediate\write18{texcount -inc -incbib 
-sum main.tex > /tmp/wordcount.tex}

\newcommand{\rew}[1]{{\color{black}{#1}}\xspace}
\newcommand{\minrew}[1]{{\color{black}{#1}}\xspace}

\begin{document}
\title{Distributed Downlink Precoding and Equalization in Satellite Swarms}

\author{
Maik Röper,~\IEEEmembership{Graduate Student Member, ~IEEE,}
	Bho Matthiesen,~\IEEEmembership{Member, ~IEEE,} 
	Dirk Wübben,~\IEEEmembership{Senior Member, ~IEEE,}
	Petar Popovski,~\IEEEmembership{Fellow, ~IEEE,} and
	Armin Dekorsy,~\IEEEmembership{Senior Member, ~IEEE}
		\thanks{
		\minrew{This article was presented in part} at 2022 IEEE Wireless Communications and Networking Conference \cite{wcnc2022}.
	}%
	\thanks{M.~R\"oper, B.~Matthiesen, D.~Wübben and A.~Dekorsy are with the Gauss-Olbers Center, c/o University of Bremen, Dept. of Communications Engineering, 28359 Bremen, Germany (email: \{roeper, matthiesen, wuebben, dekorsy\}@ant.uni-bremen.de).
	P. Popovski is with Aalborg University, Department of Electronic Systems, 9220 Aalborg, Denmark (email: petarp@es.aau.dk).
	B.~Matthiesen and P.~Popovski are also with University of Bremen, U Bremen Excellence Chair, Dept.\ of Communications Engineering, 28359 Bremen, Germany.}
\thanks{This work is supported in part by the German Federal Ministry of Education and Research (BMBF) within the project Open6GHub under grant number 16KISK016 and by the German Research Foundation (DFG) under grant EXC 2077 (University Allowance).}
}

\maketitle

\begin{abstract}
	In this paper, we propose a novel approach for downlink transmission from a satellite swarm towards a ground station (GS). These swarms have the benefit of much higher spatial separation in the transmit antennas than traditional satellites with antenna arrays, promising a massive increase in spectral efficiency. The resulting precoder and equalizer have very low demands on computational complexity, inter-satellite coordination and channel estimation. This is achieved by taking knowledge about the geometry between satellites and GS into account. For precoding, each satellite only requires its \minrew{angles of departure (AoDs)} towards the GS and it turns out that almost optimal rates can be achieved if the satellites transmit independent data streams. For the equalizer, the GS requires only knowledge about the angles of arrival (AoAs) from all satellites. Furthermore, we show that, by choosing a proper inter-satellite distance, the proposed low-complexity approach achieves the theoretical upper bound in terms of data rate. \rew{This optimal inter-satellite distance is obtained analytically under simplifying assumption and provides a heuristic for practical scenarios.} Furthermore, a novel approach to increase the robustness of the proposed precoder and equalizer against imperfect AoD and AoA knowledge is proposed by exploiting the statistics of the estimation error. 
\end{abstract}

\section{Introduction}
Integrating \acp{ntn} into terrestrial communication systems is an important step towards truly ubiquitous connectivity \cite{3GPPTR22.822,Kodheli.etal.2021}. An essential building block are small satellites in \ac{leo} that are currently deployed in private sector mega constellations \cite{Portillo.Cameron.Crawley.2019,Di.Song.Li.Poor.2019,Leyva-Mayorga2020}. Their main benefits are much lower propagation delays and deployment costs due to the \ac{leo} when compared to more traditional high-throughput satellites \cite{Zheng.Chatzinotas.Ottersten.2012,Joroughi.Vazquez.Perez-Neira.2016,Perez-Neira.Vazquez.Shankar.Malekli-Chatzinotas.2019} in \ac{meo} and \ac{geo}. While current systems focus on connecting \acp{rx} to a single satellite, combining several low cost satellites in swarms leads to increased flexibility and scalability \cite{Verhoeven.Bentum.Monna.Rotteveel.Guo.2011}.
\rew{Satellite swarms are groups of small satellite flying quite close together and often acting as a single entity. In contrast to a constellation, their purpose is not to provide global coverage. A constellation might consists of several satellite swarms. The inter-satellite distances in satellite swarms range from less than a kilometer to several hundreds of kilometer \cite{Radhakrishnan.etal.2016, Guo-Ping.2018}. In this paper, an approach to determine an appropriate inter-satellite distance for high-throughput communication is presented. Indeed,}
the joint transmission of multiple satellites forming large virtual antenna arrays promises tremendous spectral efficiency gains solely due to the increased spatial separation of antennas \cite{Budianu.Meijernik.Bentum2015,Richter.Bergel.Noam.Yair.2020}. However, the straightforward implementation of the optimal precoder requires up-to-date \ac{csi} and timely inter-satellite coordination. 
This is infeasible due to very short channel coherence times resulting from high orbital velocities in combination with comparably large propagation delays, both in ground-to-satellite and in inter-satellite links.

In this paper, we consider a satellite swarm that jointly transmits towards a \ac{rx} equipped with an antenna array, and develop a precoding approach that overcomes these obstacles by exploiting positional information about the satellites and \rew{the} \ac{rx}. In contrast to full \ac{csi}, this information is either available due to the inherent determinism of orbital mechanics or is estimated easily, even in \ac{fdd} systems, as it is based on geometric relations \cite{Im.etal.2013,Lin.Gao.Jin.Li.2017,You.Li.Wang.Gao.Xia.Ottersten.2020}.
This leads to an approximate channel model inspired by \cite{Schwarz.Delamotte.Storek.Knoop2019}, which is employed to derive a beamspace \ac{mimo} \cite{Lin.Gao.Jin.Li.2017,Ahmed2018} based distributed linear precoder and equalizer. This precoder requires only, at each satellite, relative positional knowledge between itself and the \ac{rx}. It achieves close to optimal spectral efficiency, although it only requires modest inter-satellite coordination prior to transmission instead of time-critical (and, therefore, infeasible) live inter-satellite communication during downlink transmission. It also has very low computational complexity.
We also extend the results to imperfect position knowledge and propose robust precoder and equalizer designs that exhibit considerable gain over heuristic solutions.
Similarly, the equalizer only needs \ac{aoa} information for the satellites and, given proper design of the satellite swarm, shows nearly optimal performance in terms of throughput.


In \cite{Budianu.Meijernik.Bentum2015}, the \ac{dl} from a satellite swarm with more than 50 nano-satellites towards a single antenna \acf{rx} is studied. It is shown that, if the signals of all satellites add up in phase at the \ac{rx}, a high array gain is achieved.
Communication between multiple satellites and a \ac{rx} with multiple antennas is studied in \cite{Yamashita.Kobayashi.Ueba.Umehira.2005}, where an iterative interference cancellation algorithm is considered to deal with the large spatial correlation between two close \ac{geo} satellites.
Furthermore, in \cite{Goto.Shibayama.Yamashita.Yamazato.2018} and \cite{Liolis.Panagopoulos.Cottis.2007}, the capacity of multi-satellite systems is studied.
In \cite{Roeper.Dekorsy.2019}, a distributed precoding algorithm for a multi-user \ac{dl} scenario that requires information exchange between the satellites is proposed based on the \ac{mmse} criterion.
In \cite{Richter.Bergel.Noam.Yair.2020}, a \ac{zf} equalizer at the ground terminal is proposed to simultaneously receive transmission from two satellites. \rew{The authors also numerically evaluate the impact of the inter-satellite distance with respect to (w.r.t.) the outage probability for the \ac{sinr}. Instead, we analyze, among other things, the impact of the inter-satellite distance on the ergodic rate for an arbitrary number of satellites. We also derive the optimal inter-satellite distance analytically for special cases and show that it serves as a good heuristic in all other cases.}
In \cite{You.Li.Wang.Gao.Xia.Ottersten.2020,Guo.Lu.Gao.Xia.2021,Lin.Lin.Champagne.Zhu.AlDhahir.2020}, beamspace \ac{mimo} is adapted for ground to satellite communications, focusing on scenarios involving a single satellite. In \cite{You.Li.Wang.Gao.Xia.Ottersten.2020}, a precoder for \ac{dl} transmission and an equalizer for receiving in the \ac{ul} for a single satellite communicating with  multiple devices on ground is proposed. By exploiting perfect knowledge about the geometric relations and the long term channel statistics, a similar rate could be achieved compared to the case with perfect \ac{csi} in the given scenario.

The design of robust precoders and equalizers that 
can improve the performance in presence of channel estimation errors is a well investigated subject for terrestrial scenarios \cite{book:Dietrich.2008}.
The most common assumption is that the \ac{csi} is disturbed by an additive estimation error \cite{Abrardo.Fodor.Moretti.Telek.2019,Guo.Levy.2006}.
In \cite{Mi.Zhang.Muhaidat.Tafazolli.2017,Bazzi.Xu.2016}, an extended error model for the precoder design is investigated by taking non-ideal hardware into account, resulting in an multiplicative error model.
However, such error models are invalid if the transceivers are designed based on the \acp{aoa} and \acp{aod}. Instead, this leads to a multiplicative error model with high correlation among the different antennas.
In \cite{Sun.etal.2019}, an equalizer is proposed to completely cancel interference from a broader range of possible \acp{aoa}.
Another common heuristic approach is given by the so-called diagonal loading method, where a scaled identity matrix is added to the imperfectly estimated autocorrelation matrix in order to increase robustness \cite{Carlson1988}.
Further heuristic robust equalizers are presented in \cite{Bell.Ephraim.VanTrees.2000,Mao.etal.2015}, where the diagonal loading approach is improved by taking the statistics of the estimation error into account.
The approach most similar to our proposed robust equalizer is given in \cite{Li.Zhang.Ge.Xue.2017}, where an unwanted jamming signal is suppressed while the desired signal may have its origin in a large geographical region.
However, the scenario in \cite{Li.Zhang.Ge.Xue.2017} leads to different performance metrics and, hence, different equalization problems and approaches. \rew{Furthermore, in \cite{Sun.etal.2019, Carlson1988, Bell.Ephraim.VanTrees.2000, Mao.etal.2015, Li.Zhang.Ge.Xue.2017} only 2D scenarios are investigated, i.e., there exists only one \ac{aoa} and \ac{aod} between each transmitter and receiver node. In this paper, we consider an extended 3D model with both, elevation and azimuth angle.}
In addition, we also consider arbitrarily distributed  estimation errors with known \ac{pdf}, whereas only uniformly distributed errors are considered in \cite{Bell.Ephraim.VanTrees.2000,Mao.etal.2015,Li.Zhang.Ge.Xue.2017}. This generalization complicates the analysis considerably and requires a different mathematical approach than in \cite{Bell.Ephraim.VanTrees.2000,Mao.etal.2015,Li.Zhang.Ge.Xue.2017}.

In this paper, the \ac{dl} from a satellite swarm towards a \ac{rx} is investigated. The focus is on the proper design of the distributed precoder and the corresponding equalizer, based on position knowledge, as well as the swarm layout.
In particular, the main contributions are:
\begin{itemize}
	\item A low complexity distributed precoder for satellite swarms is proposed, where the satellites transmit statistically independent data streams.
	Additionally, this precoder requires only local knowledge about the satellite's \minrew{\acp{aod}} towards the \ac{rx}. It is shown, that, with our approach, the capacity can be achieved under certain conditions and close to optimal data rates are achieved in most other cases. 
	\item An optimal linear equalizer in terms of the achievable rate based on the \ac{rx}'s knowledge of the \acp{aoa} from the satellites and the second order channel statistics is derived.
	\item The inter-satellite distance which maximizes the achievable rate in a satellite swarm is obtained for a \rew{simplified scenario. It is numerically shown that the derived distance provides a good heuristic, for broader scenarios, such that the achievable rate is still close to the channel capacity with the proposed linear precoder and equalizer.}
	\item A robust approach to reduce the performance degradation of our precoder and equalizer due to imperfect \ac{aod} and \ac{aoa} knowledge, respectively, is derived. This is achieved by taking statistical knowledge about the estimation error into account.
	This approach is not only novel in the context of satellite communications, but is also applicable to any scenario where the channel is determined by the \ac{aod} or \ac{aoa}, e.g., in mmWave communication \cite{Lin.Gao.Jin.Li.2017,Zhao.etal.2017} or radar \cite{Li.Zhang.Ge.Xue.2017}.
\end{itemize}

The system model and fundamental results serving as benchmark are introduced in Section~\ref{sec:system}. Then, in Section~\ref{sec:perCSI}, the proposed precoder and equalizer are developed under the assumption of perfect position knowledge. 
\rew{In Section~\ref{sec:impCSI}, they are extended for imperfect position knowledge.
The optimal distance between the satellites in a swarm is analyzed in Section~\ref{sec:d_sat}. This also gives further insight about the optimality of the proposed precoder.}
The proposed methods are evaluated numerically in Section~\ref{sec:simulation} and Section~\ref{sec:conclusion} concludes the paper.

\textit{Notation:}
Column vectors are denoted by bold lowercase letters $\vek{q}$, while bold uppercase letters denote  matrices $\vek{Q}$. Non-bold symbols denote scalar values $q,Q$. The $j$th element of a vector and $(j,j')$th elements of a matrix are denoted by $[\vek{q}]_j$ and $[\vek{Q}]_{j,j'}$\minrew{, respectively}. A set $\{\vek{Q}_1,\dots,\vek{Q}_J\}$ is denoted by $\{\vek{Q}_j\}_{j=1}^{J}$. Furthermore, $\diag{q_1,...,q_N}$ denotes a diagonal matrix with elements $q_1,...,q_N$ along its main diagonal. Correspondingly, block diagonal matrices are denoted by $\blkdiag{\vek{Q}_1,...,\vek{Q}_N}$. The trace, transpose and  conjugate transpose of a matrix $\vek{Q}$ are denoted by $\tr{\vek{Q}}$, $\vek{Q}^T$ and $\vek{Q}^H$, respectively.
Furthermore, $|q|$ and $|\vek{Q}|$ are the absolute value of the scalar $q$ and determinant of matrix $\vek{Q}$, respectively. The $\ell_2$-norm of a vector $\vek{q}$ is denoted as $\Vert\vek{q}\Vert_2$.
Additionally, $\nabla_{\vek{Q}}f(\vek{Q}_0)$ denotes the derivative of the function $f$ w.r.t. $\vek{Q}$ and evaluated at $\vek{Q}_0$.
Finally, \rew{$\E_{\{\vek{Q}_j\}_{j=1}^{J} }\left\{\cdot\right\}$ denotes the expected value w.r.t. to the random variables $\vek{Q}_1,\dots,\vek{Q}_J$}, $\otimes$ the Kronecker product between two matrices, $\vekI_N$ is the identity matrix of dimension $N\times N$ and $\nullvec_{N\times N'}$ is the all zero matrix of dimension $N\times N'$. 

\section{System Model and Performance Bounds}\label{sec:system}

\subsection{System Setup}
\begin{figure}
	\centering
			\includegraphics[height=2.3in]{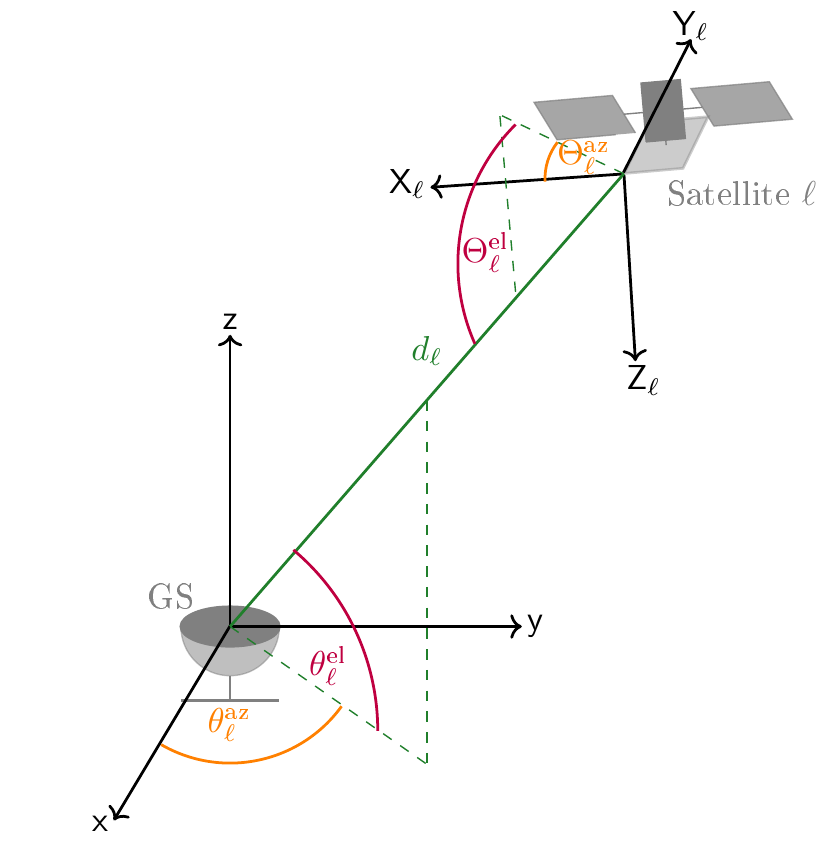}
			\caption{\rew{Visualization of local coordinate frames.}}
			\label{fig:3D_setup}
\end{figure}

Consider a swarm of $\NS$ satellites \rew{which jointly communicate to a common \ac{rx}. For each involved node, i.e., each of the $\NS$ satellites and the \ac{rx}, we can define a local coordinate reference system, as depicted in Fig. \ref{fig:3D_setup}.
The $\mathsf{xy}$-plane of each local coordinate frame is aligned with the corresponding planar antenna array. Thus, the rotation of each node, i.e., satellite or \ac{rx}, leads to a rotation of the corresponding local coordinate frame.
Let $d_\ell$ be the distance between satellite $\ell$ and the \ac{rx}, the position of that satellite in the \ac{rx}-centered coordinate system, i.e., the relative position of the satellite w.r.t. the \ac{rx}, is specified by the triplet $(d_\ell, \aoal^\text{el}, \aoal^\text{az})$, where $\aoal^\text{el}$ and $\aoal^\text{az}$ denote the elevation and azimuth angle in the \ac{rx}-centered coordinate frame, respectively. Correspondingly, the position of the \ac{rx} in the $\ell$th satellite coordinate frame is given by the triplet $(d_\ell, \aodl^\text{el}, \aodl^\text{az})$,  where $\aodl^\text{el}$ and $\aodl^\text{az}$ denote the elevation and azimuth angle in the $\ell$th satellite coordinate frame, respectively. 
Given that the \ac{rx} is located on a fixed position on the Earth and the satellites move on predefined orbits, the \acp{aoa} $\aoal^\text{el}$ and $\aoal^\text{az}$ can be extracted from the positions of satellite $\ell$ and the \ac{rx}.} 

\subsection{Generalized Communication Model} \label{sec:system_trans}
\rew{Satellite $\ell$ is equipped with an arbitrary planar antenna array consisting of $\Nt$ antennas and the \ac{rx} uses an array with $\Nr$ antennas.}
The satellites \rew{jointly} transmit a \rew{single} common message known a~priori at all satellites by encoding it in $M$ independent data streams $\veks\in\C^M$ using \ac{iid} unit variance Gaussian codebooks. Partitioning the common message into data streams requires inter-satellite coordination prior to transmission.
Satellite~$\ell$ employs linear precoding, with its local precoding matrix $\PC_\ell\in\C^{\Nt\times M}$, to transmit the signal $\vekx_\ell = \PC_\ell\veks\in\C^{\Nt}$. 
In a satellite swarm, all satellites are usually of the same type \cite{Verhoeven.Bentum.Monna.Rotteveel.Guo.2011} and, thus, we assume that all satellites have the same average transmit power constraint $\rho$, i.e.,
	$\tr{\PC_\ell\PC_\ell^H} \leq \rho$ for all $\ell=1,...,\NS.$
\rew{Furthermore, the signal  received at the \ac{rx}} is
\begin{align}\label{eq:y_sum}
    \veky = \sum_{\ell=1}^{\NS} \vekHsat\vekx_{\ell} + \vekn
\end{align}
where $\vekn$ is \ac{iid} circularly-symmetric complex white Gaussian noise with power $\sigma_\mathsf{n}^2$ and $\vekHsat\in\C^{\Nr\times\Nt}$ is the local channel matrix from satellite $\ell$ to the \ac{rx}. Due to the collaborative transmission, this is effectively a point-to-point channel. In particular, let $\vekH=\left[\vekH_1, ... ,\vekH_{\NS} \right]$ be the composite channel matrix and $\vekx=\left[\vekx_1^T, ... ,\vekx_{\NS}^T \right]^T$ the composite transmit signal. Then, we can write \eqref{eq:y_sum} as
    $\veky = \vekH \vekx + \vekn\,.$

\rew{Assuming joint precoding over all satellites, the channel capacity is upper bounded by} \cite{Telatar.1999}
\begin{align}\label{eq:rate_p2p}
	R_{\text{opt}}  
	&= \max_{\tr{\PC\PC^H} \leq \NS \rho} \log_2\left\vert \vekI_{\Nr} + \frac{1}{\sigma_\mathsf{n}^{2}} \vekH\PC\PC^H\vekH^H \right\vert,
\end{align}
where $\PC=[\PC_1^T,...,\PC_{\NS}^T]^T\in\C^{\NTx\times M}$ is the joint precoding matrix with $\NTx = \NS \Nt$ transmit antennas in total. Furthermore, we assume $M=\rank{\vekH}$. The maximum in \eqref{eq:rate_p2p} is achieved for
\begin{align}\label{eq:pc_opt}
	\PC_\text{opt} = \vekV\vekP^{\frac{1}{2}},
\end{align}
where the columns of $\vekV$ are the $M$ right singular vectors of $\vekH$, corresponding to the $M\le \NTx$ non-zero singular values,
and $\vekP=\diag{p_1,...,p_M}$ is the optimal transmit power allocation obtained from the waterfilling algorithm such that  $\sum_\mu p_\mu = \NS \rho$, with  $p_\mu$ being the transmit power of the $\mu$th stream \cite{Telatar.1999}.
Combining \eqref{eq:rate_p2p} and \eqref{eq:pc_opt}, we obtain 
\begin{align}\label{eq:rate_opt}
	R_{\text{opt}} = \sum_{\mu=1}^{M} \log_2\left(1 + \lambda_\mu\frac{p_\mu}{\nvar} \right),
\end{align}
where $\lambda_\mu$ is the $\mu$th eigenvalue of $\vekH\vekH^H$ \cite{Telatar.1999,Tse2005}. 
\rew{Note that this \ac{svd}-based precoder design incorporates several assumptions that renders it infeasible for satellite swarms. First of all, an accurate estimate of $\vekH$ is required. Due to the fast ground speeds of \ac{leo} satellites, the channel coherence time is rather short. Combined with the long round trip times between satellites and \ac{rx}, obtaining this estimate with conventional methods appears to be impossible. Further, assuming a timely and accurate estimate of $\vekH_\ell$ exists at satellite $\ell$, this estimate would have to be shared with all other satellites within the swarm, leading to additional delay and, thus, ageing of the channel estimate.
Finally, the bound in \eqref{eq:rate_p2p} has a relaxed sum power constraint over all satellites that might lead to the violation of the power constraints of individual satellites.
However, while being of little practical relevance,} it gives the theoretical upper bound for the achievable rate \rew{that can serve as benchmark. In the following, we design a precoder and an equalizer that require only knowledge of rather long-term fading statistics and positional knowledge of the satellites and the \ac{rx}, together with synchronized clocks between satellites for coherent transmission.}

\section{Geometry Based DL Transmission}\label{sec:perCSI}
\rew{In this section, we exploit positional information about the satellites and the \ac{rx} to estimate the dominant large-scale components of the channel matrix $\vekH$. This results in a geometrical channel model that is subsequently employed to design a distributed linear precoder together with a linear equalization scheme at the \ac{rx}. This approach does not require any inter-satellite coordination beyond clock synchronization and is not subject to the limitations of conventional \ac{csi} acquisition resulting from the short channel coherence time and long transmission distances.}

\subsection{Geometrical Channel Approximation} \label{sec:cm}
\rew{Observe that the channels from the antennas of a single satellite to the \ac{rx} are highly correlated \cite{Schwarz.Delamotte.Storek.Knoop2019} and, thus, there are only $M\le \NS\le \NTx$ singular values larger than zero. Further, assume that the communication between satellites and the \ac{rx} takes place under \ac{los} conditions, which is a very typical scenario in satellite communications. Then,
the channel matrix $\vekH$ is exclusively determined by the distances 
between transmit and receive antennas as well as atmospheric effects \cite{3GPP.TR.38.811,Storek.Hofmann.Knopp.2015}.}
\rew{Moreover, due to the \ac{los} connection and the large distance between satellites and \ac{rx}, 
the channels of the antennas from satellite $\ell$ to the ground station are subject to approximately the same atmospheric effects \cite{Storek.Hofmann.Knopp.2015}.} Thus, it is reasonable to assume that the entries in $\vekH_\ell$ have equal magnitude and differ only in their phase. 

\begin{figure}
		\begin{subfigure}{0.47\columnwidth}
			\includegraphics[height=1.75in]{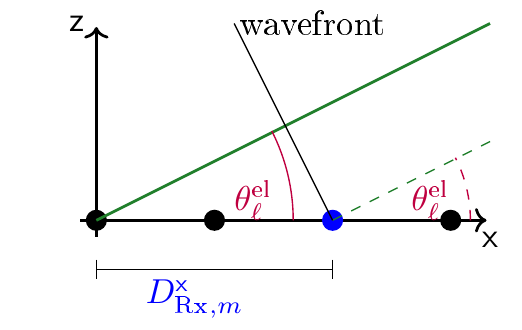}
			\caption{\rew{Side view of the \ac{rx}'s \ac{ura}.}}
		\end{subfigure}
		\begin{subfigure}{0.47\columnwidth}
				\includegraphics[height=1.75in]{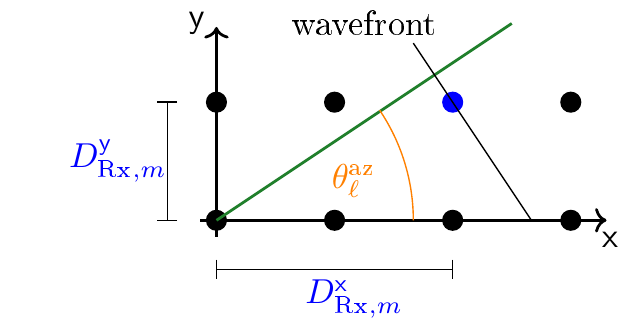}
				\caption{\rew{Top view of the \ac{rx}'s \ac{ura}.}}
			\end{subfigure}	
	\caption{\rew{Visualization of the \acp{aoa} for a wavefront arriving at the \ac{rx} with a \ac{ura}.}}
	\label{fig:3D_aoa}
\end{figure}

\rew{Let $(D_{\text{Rx},m}^{\mathsf{x}},D_{\text{Rx},m}^{\mathsf{y}},0)$ be the Cartesian coordinates of the $m$th antenna at the \ac{rx}, given in the \ac{rx}-centered coordinate frame, as depicted in Fig. \ref{fig:3D_aoa} for an exemplary \ac{ura}. Then, due to the large distance between the satellites and the \ac{rx}, the electromagnetic wave radiated by satellite $\ell$ arrives approximately as a plane wave \cite{3GPP.TR.38.811,You.Li.Wang.Gao.Xia.Ottersten.2020} and the phase rotation due to the transmission channel observed at the $m$th \ac{rx}-antenna is approximately
\begin{subequations}
	\begin{align}
		\phi_m^\ell &\approx \nu\left(D_{\text{Rx},m}^{\mathsf{x}} \cos\left(\aoa_\ell^{\text{el}}\right) \cos\left(\aoa_{\ell}^{\text{az}}\right) + D_{\text{Rx},m}^{\mathsf{y}} \cos\left(\aoa_\ell^{\text{el}}\right) \sin\left(\aoa_{\ell}^{\text{az}}\right) \right) + \phi_0^\ell\\
		&= \nu \left(D_{\text{Rx},m}^{\mathsf{x}} \phi_{\ell}^{\mathsf{x}} + D_{\text{Rx},m}^{\mathsf{y}} \phi_{\ell}^{\mathsf{y}}\right) + \phi_0^\ell, \quad m=1,...,\Nr
	\end{align}
\end{subequations}
where $\phi_{\ell}^{\mathsf{x}}= \cos(\aoa_\ell^{\text{el}}) \cos(\aoa_{\ell}^{\text{az}}), \phi_{\ell}^{\mathsf{y}} = \cos(\aoa_\ell^{\text{el}}) \sin(\aoa_{\ell}^{\text{az}})$ are the space angles in $\mathsf{x}$- and $\mathsf{y}$-direction, respectively, $\phi_0^\ell$ is a phase offset, which is the same for all receive antennas, and $\nu= 2\pi \fc/c_0$ is the wavenumber.}

\rew{Likewise, let $(D_{\text{Tx},n}^{\mathsf{x},\ell},D_{\text{Tx},n}^{\mathsf{y},\ell},0)$ be the Cartesian coordinates of the $n$th antenna of satellite $\ell$, given in the $\ell$th satellite-centered coordinate frame.
Then, the signal, radiated by the $n$th antenna of satellite $\ell$ and arriving at the $m=1$st receive antenna at the \ac{rx}, is rotated by
\begin{subequations}
	\begin{align}
		\Phi_n^\ell  &\approx \nu\left(D_{\text{Tx},n}^{\mathsf{x},\ell} \cos\left(\aod_\ell^{\text{el}}\right) \cos\left(\aod_{\ell}^{\text{az}}\right) + D_{\text{Tx},n}^{\mathsf{y},\ell} \cos\left(\aod_\ell^{\text{el}}\right) \sin\left(\aod_{\ell}^{\text{az}}\right) \right) + \Phi_0^\ell \\
		&= \nu \left(D_{\text{Tx},n}^{\mathsf{x},\ell} \Phi_{\ell}^{\mathsf{x}} +  D_{\text{Tx},n}^{\mathsf{y},\ell}\Phi_{\ell}^{\mathsf{y}}\right) + \Phi_0^\ell, \quad n=1,...,\Nt
	\end{align}
\end{subequations}
where $\Phi_{\ell}^{\mathsf{x}}= \cos(\aod_\ell^{\text{el}}) \cos(\aod_{\ell}^{\text{az}}), \Phi_{\ell}^{\mathsf{y}} = \cos(\aod_\ell^{\text{el}}) \sin(\aod_{\ell}^{\text{az}})$ are the space angles in $\mathsf{x}$- and $\mathsf{y}$-direction, respectively and $\Phi_0^\ell$ is a phase offset, which is the same for all antennas at satellite $\ell$.
Note that the phase offsets $\phi_0^\ell\in[0,2\pi]$ and $\Phi_0^\ell\in[0,2\pi]$ include the distance $d_\ell$ as well as the atmospheric effects.}
Now, the approximated channel coefficient $\tilde{h}_{m,n}^\ell$ from the $n$th antenna at satellite $\ell$ to the $m$th receive antenna can be written as
\rew{
\begin{align}\label{eq:ch_appr_elements}
    \tilde{h}_{m,n}^\ell
    &=  \alpha_\ell e^{j\nu \left( D_{\text{Rx},m}^{\mathsf{x}} \phi_{\ell}^{\mathsf{x}} + D_{\text{Rx},m}^{\mathsf{y}} \phi_{\ell}^{\mathsf{y}} + D_{\text{Tx},n}^{\mathsf{x},\ell} \Phi_{\ell}^{\mathsf{x}} +  D_{\text{Tx},n}^{\mathsf{y},\ell}\Phi_{\ell}^{\mathsf{y}} \right)}
\end{align}
where $\alpha_\ell=|\alpha_\ell|e^{j(\Phi_0^\ell + \phi_0^\ell)}$ is the \ac{iid} complex-valued channel gain from satellite $\ell$ to the \ac{rx} with $\E_{\alpha_\ell}\{\alpha_\ell\}=0$ and
$\E_{\alpha_\ell}\{|\alpha_\ell|^2\}=\sigma_{\alpha_\ell}^2$, which includes the free space path loss 
and the atmospheric effects.} 
From these observations, we can define the steering vectors $\steeraoa_\ell$ and $\steeraod_\ell$. The vector $\steeraoa_\ell$ contains the phase differences at the receive antennas, while $\steeraod_\ell$ contains the phase differences from the transmit antennas at satellite $\ell$. Thus, their elements are
	\rew{\begin{align}\label{eq:steering}
		\left[\steeraoa_\ell\right]_m  = a_{\ell,m} = e^{j\nu\left(D_{\text{Rx},m}^{\mathsf{x}} \phi_{\ell}^{\mathsf{x}} + D_{\text{Rx},m}^{\mathsf{y}} \phi_{\ell}^{\mathsf{y}} \right)} ,\qquad
		\left[\steeraod_\ell\right]_n  = b_{\ell,n} = e^{-j \nu\left(D_{\text{Tx},m}^{\mathsf{x},\ell} \Phi_{\ell}^{\mathsf{x}} + D_{\text{Tx},m}^{\mathsf{y},\ell} \Phi_{\ell}^{\mathsf{y}} \right)}
	\end{align}}
and the approximate channel matrix becomes
\begin{align}\label{eq:ch_appr_sat}
	\chest_\ell = \alpha_\ell\veka_\ell\vekb_\ell^H \approx \chtx{\ell}\,.
\end{align}
Further, we define $\Steeraoa=[\steeraoa_1,...,\steeraoa_{\NS}]$.
Observe that $\chest_\ell$ has rank one if $\alpha_\ell\ne 0$, for all $\ell$. 
Then, due to $\Nr\ge\NS$ and the satellites having different positions in the orbital plane, i.e., $\aoa_i\neq\aoa_\ell$, for all $i\neq\ell$, the overall channel matrix $\chest=[\alpha_1\veka_1\vekb_1^H,...,\alpha_{\NS}\veka_{\NS}\vekb_{\NS}^H]$ has rank $\NS$, which allows for the parallel transmission of $M = \NS$ independent streams.
This, however, is only true for the approximated channel and does not necessarily hold for the real channel. Indeed, $M$ can be less than $\NS$ for a very small inter-satellite distance and this approach stops working. The choice of a proper inter-satellite distance will be discussed further in Section~\ref{sec:d_sat}.

This geometric approximation simplifies channel estimation and tracking considerably, mainly due to two reasons: First, the angles \rew{$\aoal^\text{el}, \aoal^\text{az}, \aodl^\text{el},\aodl^\text{az}$ and, correspondingly, the space angles $\phi_{\ell}^{\mathsf{x}}, \phi_{\ell}^{\mathsf{y}}, \Phi_{\ell}^{\mathsf{x}}, \Phi_{\ell}^{\mathsf{y}}$} are independent of the carrier frequency and, thus, can be easily estimated in \ac{fdd} systems \cite{Im.etal.2013}.  
Second, a~priori knowledge about the positions and the deterministic satellite movement can be exploited to increase estimation accuracy.
In the following, the precoder and equalizer are designed based on the steering vectors $\steeraod_\ell$ and $\steeraoa_\ell$ \minrew{\eqref{eq:steering}}, respectively.

\subsection{Precoding}\label{sec:ad_pc}
Given the geometric channel approximation $\chest$, we design a precoder $\PC_{\text{geo}}$ that maximizes the average received power given the per-satellite power constraint, i.e., the solution to
\begin{align}\label{eq:opt_pc_geo}
		\PC_{\text{geo}}\! \in \argmax_{\forall \ell: \tr{\PC_\ell^H\PC_\ell} \le \rho}\enskip  \E_{\left\{\alpha_\ell\right\}_{\ell=1}^{\NS}} \left\{\tr{\PC^H\chest^H\chest\PC}\right\} .
\end{align}
An optimal precoder with respect to \eqref{eq:opt_pc_geo} is stated in the following proposition.
\begin{proposition}\label{prop:pc}
	A precoder that maximizes the received signal power under per-satellite power constraint, 
	i.e., a solution to \eqref{eq:opt_pc_geo}, is
	\begin{align}\label{eq:pc_ad}
		\PC_\mathrm{geo} 
		= \sqrt{\frac{\rho}{\Nt}} \blkdiag{\steeraod_1, \dots ,\steeraod_{\NS}}
		= \blkdiag{\pc_{1,\mathrm{geo}}, \dots ,\pc_{\NS,\mathrm{geo}}}\,.
	\end{align}  
\end{proposition}

\begin{IEEEproof}
Define the matrices $\Steeraod=\blkdiag{\steeraod_1, \dots ,\steeraod_{\NS}}$ and $\vekSig_\alpha=\diag{\alpha_1,\dots,\alpha_{\NS}}$. Then, the channel matrix $\chest$ can be decomposed as
	$\chest = \Steeraoa\vekSig_\alpha\Steeraod^H$.
%
Note that $\Steeraoa$ and $\Steeraod$ are both deterministic. Thus, the objective function in \eqref{eq:opt_pc_geo} is equivalent to
\begin{align}
	\MoveEqLeft\E_{\left\{\alpha_\ell\right\}_{\ell=1}^{\NS}} \left\{\tr{\PC^H\chest^H\chest\PC}\right\} \notag\\
	&= \E\left\{\tr{\PC^H\Steeraod\vekSig_\alpha^H\Steeraoa^H\Steeraoa\vekSig_\alpha\Steeraod^H\PC}\right\}
	= \tr{\PC^H\Steeraod\, \E \left\{\vekSig_\alpha^H\Steeraoa^H\Steeraoa\vekSig_\alpha\right\}\Steeraod^H\PC} \notag\\
	&\rew{= \tr{\PC^H\Steeraod \Nr\E \left\{\vekSig_\alpha^H\vekSig_\alpha\right\} \Steeraod^H\PC}}
	\rew{= \Nr\sum_{\ell=1}^{\NS} \sigma_{\alpha_\ell}^2 \tr{\PC_\ell^H\steeraod_\ell\steeraod_\ell^H\PC_\ell}} .\label{eq:objective_pc_geo2}
\end{align}
Since $\steeraod_\ell\steeraod_\ell^H$ is positive semi-definite and \rew{$\sigma_{\alpha_\ell}^2$ and $\Nr$ are positive constants}, the objective function is maximized if each term in \eqref{eq:objective_pc_geo2} is maximized separately.
Hence, the optimization problem \eqref{eq:opt_pc_geo} is equivalent to
\begin{align}\label{eq:opt_pc_geo_alt}
	\min_{\PC_\ell}\enskip -\tr{\PC_\ell^H\steeraod_\ell\steeraod_\ell^H\PC_\ell} \quad\text{s.t.}\quad \tr{\PC_\ell^H\PC_\ell} \le \rho,\quad \ell=1, \dots, \NS 
\end{align}
and the corresponding Lagrangian function is
\begin{align}\label{eq:lag_pc_geo}
	\lagfunc_{\text{pc}}\left(\PC_\ell,\lagrange_\ell\right) &= -\tr{\PC_\ell^H\steeraod_\ell\steeraod_\ell^H\PC_\ell} + \lagrange_\ell\left(\tr{\PC_\ell^H\PC_\ell} - \rho\right)
\end{align}
where $\lagrange_\ell\ge 0$ is the Lagrange multiplier corresponding to the $\ell$th power constraint.

We leverage \cite[Prop.~3.3.4]{book:Bertsekas.1999} to find the solution of \eqref{eq:opt_pc_geo_alt}.
Every stationary point $\PC_{\ell}'$ satisfies
    $\nabla_{\PC}\lagfunc_{\text{pc}}\left(\PC_{\ell}',\lagrange_{\ell}'\right) = -\PC_{\ell}'^H\steeraod_\ell\steeraod_\ell^H + \lagrange_{\ell}'\PC_{\ell}'^H = \nullvec$,
for some $\lagrange_{\ell}'$.
This is  equivalent to
\begin{equation} \label{eq:eig}
    \steeraod_\ell\steeraod_\ell^H\PC_{\ell}' = \lagrange_{\ell}'\PC_{\ell}',
\end{equation}
which holds either if $\PC_{\ell}'=\nullvec_{\Nt\times\NS}$ or $\lagrange_{\ell}'$ is an eigenvalue of $\steeraod_\ell\steeraod_\ell^H$. Since, $\steeraod_\ell\steeraod_\ell^H$ is of rank one, its only eigenvalue is given by \cite[Thm.~1.2.12]{Horn1990}
\begin{align} \label{eq:omega}
	\lagrange_{\ell}' = \tr{\steeraod_\ell\steeraod_\ell^H} = \sum_{n=1}^{\Nt}e^{j0} = \Nt .
\end{align}
Consider $\PC_{\ell}'=\nullvec$. Then, 
$\lagfunc_{\text{pc}}\left(\nullvec,\lagrange_\ell\right) = -\lagrange_\ell\rho$. Since $\tr{\PC_\ell^{'H}\PC_\ell'} < \rho$, $\lagrange_{\ell}^* = 0$ \cite[Prop.~3.3.4]{book:Bertsekas.1999}. Hence, the case $\PC_{\ell}'=\nullvec$ is irrelevant when considering $\lagrange_{\ell}' = \Nt$.

Substitute \eqref{eq:eig} and \eqref{eq:omega} into \eqref{eq:lag_pc_geo}. Then, at every stationary point with $\PC_{\ell}'\neq\nullvec$,
\begin{equation}\label{eq:lag_pc_geo2}
    \lagfunc_{\text{pc}}\left(\PC_{\ell}',\lagrange_{\ell}'\right) = -\Nt \tr{\PC_{\ell}'^{H} \PC_{\ell}'} + \Nt \left(\tr{\PC_{\ell}'^{H}\PC_{\ell}'} - \rho\right) = -\Nt \rho
\end{equation}
which is the minimum value of $\lagfunc_{\text{pc}}\left(\PC_{\ell},\lagrange_{\ell}'\right)$.
Since
\begin{align}
    \lagfunc_{\text{pc}}\left(\PC_{\ell,\text{geo}},\lagrange_{\ell}'\right) &= -\tr{\sqrt{\frac{\rho}{\Nt}}\steeraod_\ell^H\steeraod_\ell\steeraod_\ell^H\sqrt{\frac{\rho}{\Nt}}\steeraod_\ell} + \Nt\left( \tr{\sqrt{\frac{\rho}{\Nt}}\steeraod_\ell^H\sqrt{\frac{\rho}{\Nt}}\steeraod_\ell} - \rho\right)  \\
    &= -\frac{\rho}{\Nt} (\steeraod_\ell^H\steeraod_\ell) (\steeraod_\ell^H\steeraod_\ell) +  \rho(\steeraod_\ell^H\steeraod_\ell) - \Nt \rho
    = -\Nt \rho
\end{align}
$\PC_{\ell,\text{geo}}$ is a global minimizer of $\lagfunc_{\text{pc}}\left(\PC_{\ell},\lagrange_{\ell}'\right)$. Further, since
\begin{equation}
    \tr{\PC_{\ell,\text{geo}}^H\PC_{\ell,\text{geo}}} = \tr{\sqrt{\frac{\rho}{\Nt}}\steeraod_\ell^H\sqrt{\frac{\rho}{\Nt}}\steeraod_\ell} = \rho
\end{equation}
$\PC_{\ell,\text{geo}}$ is a solution to \eqref{eq:opt_pc_geo_alt} \cite[Prop.~3.3.4]{book:Bertsekas.1999}.
\end{IEEEproof}

A direct consequence of the block diagonal precoding matrix $\PC_{\text{geo}}$ is that each satellite transmits a different independent stream $s_\ell$ of the whole data vector $\veks$, i.e., the transmit signal of satellite $\ell$ is 
   $ \vekx_\ell = \PC_{\ell,\text{geo}}\veks = \pc_{\ell,\text{geo}} s_\ell$.
Furthermore, the precoding vector $\pc_{\ell,\text{geo}}$ for satellite $\ell$ is independent of the precoding vectors $\pc_{i,\text{geo}}$ for the other satellites $i\ne\ell$. Instead, satellite $\ell$ only has to know its \ac{aod} $\aodl$.
This means that, due to the geometric channel approximation, the optimal solution \rew{to} \eqref{eq:opt_pc_geo} can be obtained if the data stream is split among the satellites and each satellite processes its precoding vector locally. \rew{This is remarkable in so far that cooperative \ac{dl} transmission is enabled without the requirement for inter-satellite information exchange during the transmission.}
Additionally, no costly \ac{svd} is needed for this approach.
Instead, the proposed precoding is based on manipulating only the phase at each antenna and, thus, an efficient implementation with a single RF chain per satellite is possible \cite{Lin.Gao.Jin.Li.2017}.

\subsection{Linear Equalization}\label{sec:ad_eq}
Employing the previously derived precoder $\PC_\text{geo}$, the received signal at the \ac{rx} is 
\begin{align}
    \veky = \vekH\PC_\text{geo}\veks + \vekn = \sum_{\ell=1}^{\NS} \chtx{\ell}\pc_{\ell,\text{geo}}s_\ell + \vekn \,.
\end{align}
Now, we consider linear reception at the \ac{rx} to recover the $\NS$ data streams. In particular, after linear equalization with $\EQ=[\eq_1,...,\eq_{\NS}]^H\in\C^{\NS\times\Nr}$, the $\ell$th estimated stream is 
\begin{align}
    \hat{s}_\ell = \eq_\ell^H\veky = \eq_\ell^H \vekH_\ell\pc_{\ell,\text{geo}}s_\ell + \eq_\ell^H\left(\sum_{i\ne \ell} \vekH_i\pc_{i,\text{geo}}s_i + \vekn\right) \,.
\end{align}
Assuming the use of capacity achieving point-to-point codes for all streams, the achievable rate $R_\text{lin}$ is the sum over the maximum per-stream rates \cite[\S 8.3]{Tse2005}, i.e.,
\begin{align}\label{eq:rate_sum}
	R_{\text{lin}} = \sum_{\ell=1}^{\NS}R_\ell = \sum_{\ell=1}^{\NS} \log_2\left(1 + \Gamma_\ell\right)
\end{align}
where
\begin{align}
	\Gamma_\ell &= \frac{\left\vert\eqsat^H\chtx{\ell}\pc_{\ell,\text{geo}}\right\vert^2} {\sum_{i\neq\ell} \left\vert \eqsat^H\chtx{i}\pc_{i,\text{geo}} \right\vert^2 + \nvar\eqsat^H\eqsat} \label{eq:sinr}
\end{align}
is the effective \ac{sinr} of the $\ell$th stream.
Observe that $\Gamma_\ell$ is not a function of $\eq_i$ for all $i\neq\ell$. Thus, the rate in \eqref{eq:rate_sum} is maximized by maximizing each term separately. Due to the monotonicity of the logarithm, this optimum is obtained at the maximum $\Gamma_\ell$. However, determining the optimal equalizers with respect to \eqref{eq:sinr} would require full \ac{csi} knowledge at the \ac{rx}, which is difficult to obtain \cite{You.Li.Wang.Gao.Xia.Ottersten.2020}. 
Instead, we employ the geometrical approximation $\chest$ of $\vekH$ and derive an equalizer with respect to $\Gamma_\ell|_{\vekH=\chest}$. Since $\chest_\ell$ still depends on the unknown fading factor $\alpha_{\ell}$, we optimize the mean \ac{sinr} instead of the instantaneous, i.e.,
\begin{align}\label{eq:opt_eq_geo}
    \eq_{\ell,\text{geo}} \in \argmax_{\eq_\ell}\enskip \E_{\left\{\alpha_\ell'\right\}_{\ell'=1}^{\NS}} \left\{\frac{\left\vert\eqsat^H\chest_{\ell}\pc_{\ell,\text{geo}}\right\vert^2} {\sum_{i\neq\ell} \left\vert \eqsat^H\chest_{i}\pc_{i,\text{geo}} \right\vert^2 + \nvar\eqsat^H\eqsat} \right\}\,.
\end{align}
A solution to this optimization problem is stated next.

\begin{proposition} \label{prop:eq}
    A linear equalizer $\eq_\ell$ that maximizes the effective mean \ac{sinr} of the approximated channel $\chest$, i.e., a solution to \eqref{eq:opt_eq_geo}, is
		\begin{align}\label{eq:eq_geo}
		\eq_{\ell,\mathrm{geo}} &= \left(\rew{\sum_{i}^{\NS} \sigma_{\alpha_i}^2 \veka_i \veka_i^H} + \naltvarl \vekI_{\Nr} \right)^{-1} \veka_\ell  
		=  \left(\vekA\rew{\Sigma_{|\alpha|^2}}\vekA^H + \naltvar\vekI_{\Nr} \right)^{-1} \steeraoa_\ell
		\end{align}
    where \rew{$\naltvar=\frac{\nvar}{\Nt
    \rho}$ is the inverse \ac{snr} and $\Sigma_{|\alpha|^2}=\diag{\sigma_{\alpha_1}^2, ... , \sigma_{\alpha_{\NS}}^2}$.} 
\end{proposition}

\begin{IEEEproof}
     The objective function of \eqref{eq:opt_eq_geo} is equivalent to
\rew{\begin{subequations}\label{eq:eqproof1}
     \begin{align}
   		\E_{\left\{\alpha_\ell'\right\}_{\ell'=1}^{\NS}} \left\{\Gamma_\ell \right\}|_{\vekH=\chest} &= \E_{\left\{\alpha_\ell'\right\}_{\ell'=1}^{\NS}} \left\{\frac{\Nt\rho \left\vert \alpha_\ell \eqsat^H \veka_\ell\right\vert^2} {\sum_{i\neq\ell} \Nt\rho \left\vert \alpha_i \eqsat^H \veka_i \right\vert^2 + \nvar\eqsat^H\eqsat} \right\} \\
		 &= \frac{\E_{\alpha_{\ell}} \left\{\left\vert \alpha_\ell\right\vert^2\right\} \left\vert\eqsat^H \veka_\ell\right\vert^2} {\sum_{i\neq\ell} \E_{\alpha_i} \left\{\left\vert \alpha_i\right\vert^2 \right\} \left\vert \eqsat^H \veka_i \right\vert^2 + \frac{\nvar}{\Nt\rho}\eqsat^H\eqsat} \\
    	%
    	&= \frac{ \sigma_{\alpha_\ell}^2 \eqsat^H \veka_\ell \veka_\ell^H \eqsat} {\eqsat^H \left(\sum_{i\neq\ell} \sigma_{\alpha_i}^2 \veka_i \veka_i^H + \naltvar \vekI_{\Nr} \right) \eqsat}.\label{eq:eqproof2}
    \end{align} 
    \end{subequations}}

    Observe that this is a generalized Rayleigh quotient.    
    By virtue of Lemma~\ref{lem:rayleigh} in Appendix~\ref{sec:rayleigh}, \eqref{eq:eqproof2} is maximized if $\eqsat$ is  an eigenvector corresponding to the maximum eigenvalue of \\ $\left(\sum_{i\neq\ell} \sigma_{\alpha_i}^2 \rew{\veka_i \veka_i^H + \naltvar \vekI_{\Nr}} \right)^{-1} \veka_\ell \veka_\ell^H.$ It is established in \cite[\S 3]{Patcharamaneepakorn.Armour.Doufexi.2012} that \eqref{eq:eq_geo} is such an eigenvector.
\end{IEEEproof}

Stacking the individual equalizer $\{\eq_i\}_{i=1}^{\NS}$ gives the overall equalizer matrix
\rew{\begin{align}
		\EQ_\text{geo} 
		= \vekA^H \left(\vekA\Sigma_{|\alpha|^2}\vekA^H + \naltvar\vekI_{\Nr} \right)^{-1} 
		= \left(\Sigma_{|\alpha|^2}\vekA^H\vekA + \naltvar\vekI_{\NS} \right)^{-1}\vekA^H \,.
	\end{align}}
Note that the proposed closed-form solution only requires the knowledge of the \acp{aoa} $\{\aoa_i\}_{i=1}^{\NS}$ from all satellites as well as the \ac{snr} \rew{$\Nt\rho/\nvar$ and the path losses $\{\sigma_{\alpha_i}^2\}_{i=1}^{\NS}$}. 
Further note, that any $\beta \eq_{\ell,\text{geo}}$, with $\beta\ne 0$, is a valid solution for \eqref{eq:opt_eq_geo}. Common choices are $\beta=1$ \cite{You.Li.Wang.Gao.Xia.Ottersten.2020}, or   $\beta=1/\vert\eq_{\ell,\text{geo}}^H\vekH_\ell\pc_\ell\vert$ \cite{Bell.Ephraim.VanTrees.2000,Li.Zhang.Ge.Xue.2017}.

\section{Imperfect Position Knowledge}\label{sec:impCSI}
So far, perfect knowledge about \rew{the angles $\aoal^{\text{el}}$, $\aoal^{\text{az}}$, $\aodl^{\text{el}}$ and $\aodl^{\text{az}}$ for satellite $\ell$} has been assumed. Obvioulsy, in practical systems, only an estimate \rew{of these angles} is available.
In this section, the proposed precoder and equalizer are extended to be less sensitive against \rew{such estimation errors}.

Usually, \rew{the \acp{aod} and \acp{aoa}} are estimated by measuring the phase difference between the antennas, \rew{ i.e., by estimating the space angles $\phi_{\ell}^{\mathsf{x}}, \phi_{\ell}^{\mathsf{y}}, \Phi_{\ell}^{\mathsf{x}}, \Phi_{\ell}^{\mathsf{y}}$ first}\cite{Im.etal.2013,Lin.Gao.Jin.Li.2017}. Due to phase noise, this can only be estimated imperfectly and, therefore, the following error model is considered:\footnote{Note that, in principle, it is also possible to directly track the \ac{aoa} of an incoming wave \cite{Zhao.etal.2017}. However, this is computational more complex due to its high non-linearity and not necessary in the considered scenario.}
\rew{\begin{subequations} \label{eq:est_error_angles}
 \begin{align}
    \hat{\phi}_{\ell}^{\mathsf{x}} 
    &= \phi_{\ell}^{\mathsf{x}} + \erraoa_\ell^{\mathsf{x}}, \qquad
    \hat{\phi}_{\ell}^{\mathsf{y}} 
    = \phi_{\ell}^{\mathsf{y}} + \erraoa_\ell^{\mathsf{y}} \\
	\hat{\Phi}_{\ell}^{\mathsf{x}} 
	&= \Phi_{\ell}^{\mathsf{x}} + \erraod_\ell^{\mathsf{x}}, \qquad 
	\hat{\Phi}_{\ell}^{\mathsf{y}} 
	= \Phi_{\ell}^{\mathsf{y}} + \erraod_\ell^{\mathsf{y}}
\end{align}   
\end{subequations}
where $\erraoa_\ell^{\mathsf{x}}$, $\erraoa_\ell^{\mathsf{y}}$, $\erraod_\ell^{\mathsf{x}}$, $\erraod_\ell^{\mathsf{y}}$ are zero mean and statistically independent random variables. Furthermore, $\erraoa_\ell^{\mathsf{x}}$ and $\erraoa_\ell^{\mathsf{y}}$, and $\erraod_\ell^{\mathsf{x}}$ and $\erraod_\ell^{\mathsf{y}}$ are following the same distribution, respectively.}
It is further assumed that, for each satellite $\ell$, these estimation errors follow the same distribution.
Correspondingly, one can define the estimated steering vectors $\steeraoaest_\ell$ and $\steeraodest_\ell$ by substituting the true angle in \eqref{eq:steering} with the estimated one. Thus, the elements for satellite $\ell$ are \rew{
	\begin{align}
		[\steeraoaest_\ell]_m = \hat{a}_{\ell,m} = e^{j\nu\left(D_{\text{Rx},m}^{\mathsf{x}} \hat{\phi}_{\ell}^{\mathsf{x}} + D_{\text{Rx},m}^{\mathsf{y}} \hat{\phi}_{\ell}^{\mathsf{y}} \right)}, \qquad 
		[\steeraodest_\ell]_n = \hat{b}_{\ell,n}  = e^{j\nu\left(D_{\text{Tx},n}^{\mathsf{x}} \hat{\Phi}_{\ell}^{\mathsf{x}} + D_{\text{Tx},n}^{\mathsf{y}} \hat{\Phi}_{\ell}^{\mathsf{y}} \right)}
	\end{align}
}
A common assumption in the literature is that these estimation errors are uniformly distributed \cite{Lin.Lin.Champagne.Zhu.AlDhahir.2020,Bazzi.Xu.2016,Sun.etal.2019,Mao.etal.2015,Li.Zhang.Ge.Xue.2017}. However, in many applications, such estimation errors are better modeled by a Gaussian distribution \cite{Mi.Zhang.Muhaidat.Tafazolli.2017}. Therefore, we derive the precoder and equalizer for arbitrarily distributed estimation errors and show how to determine them for both, uniform and Gaussian distributed cases.
For a better understanding, we first show the statistic relation between the space angles and the corresponding steering vectors, which is related to the \ac{cf}. Following that, we derive the precoder and equalizer, utilizing the compact notation given with the \ac{cf}. To the best of the authors' knowledge, this approach is novel within the context of analyzing angular estimation errors.

\subsection{Characteristic Function of Estimation Error}\label{sec:cf}
\rew{From \eqref{eq:est_error_angles}, it follows that the true space angles $\phi_{\ell}^{\mathsf{x}}, \phi_{\ell}^{\mathsf{y}}, \Phi_{\ell}^{\mathsf{x}}$ and  $\Phi_{\ell}^{\mathsf{y}}$ are random variable with mean $\hat{\phi}_{\ell}^{\mathsf{x}}, \hat{\phi}_{\ell}^{\mathsf{y}}, \hat{\Phi}_{\ell}^{\mathsf{x}}$ and $\hat{\Phi}_{\ell}^{\mathsf{y}}$, respectively. The shape of their \ac{pdf} is determined by the \ac{pdf} of $\erraoa_\ell^{\mathsf{x}}$, $\erraoa_\ell^{\mathsf{y}}$, $\erraod_\ell^{\mathsf{x}}$, $\erraod_\ell^{\mathsf{y}}$, respectively.}
Let $\pdfaoa(\erraoa_\ell^{\mathsf{x}})$ be the \ac{pdf} of the estimation error $\erraoa_\ell^{\mathsf{x}}$. Then, its \ac{cf} $\cfaoa(t)$ is defined as \cite{Kobayashi.Mark.Turin.2011}
\begin{align}\label{eq:cf_aoa}
    \cfaoa(t) = \E_{\erraoa_\ell^{\mathsf{x}}} \left\{e^{jt\erraoa_\ell^{\mathsf{x}}} \right\} = \int_{-\infty}^{\infty} \pdfaoa(\erraoa_\ell) \, e^{jt\erraoa_\ell^{\mathsf{x}}} \mathrm{d}\erraoa_\ell^{\mathsf{x}} \,.
\end{align}
Hence, the expected value of the $m$th element $a_{\ell,m}$ of the steering vector $\steeraoa_\ell$ can be written as
\rew{
\begin{subequations}\label{eq:expec_a}
   \begin{align}
    \E_{\erraoa_\ell^{\mathsf{x}},\erraoa_\ell^{\mathsf{y}}} \left\{a_{\ell,m}\right\} &= \E_{\erraoa_\ell^{\mathsf{x}},\erraoa_\ell^{\mathsf{y}}} \left\{e^{j\nu\left(D_{\text{Rx},m}^{\mathsf{x}} \phi_{\ell}^{\mathsf{x}} + D_{\text{Rx},m}^{\mathsf{y}} \phi_{\ell}^{\mathsf{y}} \right)} \right\} \\
    &= e^{j\nu\left(D_{\text{Rx},m}^{\mathsf{x}} \hat{\phi}_{\ell}^{\mathsf{x}} + D_{\text{Rx},m}^{\mathsf{y}} \hat{\phi}_{\ell}^{\mathsf{y}} \right)}
    \E_{\erraoa_\ell^{\mathsf{x}}} \left\{e^{-j\nu D_{\text{Rx},m}^{\mathsf{x}}\erraoa_\ell^{\mathsf{x}} } \right\} 
	\E_{\erraoa_\ell^{\mathsf{y}}} \left\{e^{-j\nu D_{\text{Rx},m}^{\mathsf{y}}\erraoa_\ell^{\mathsf{y}} } \right\} \\
    &= e^{j\nu\left(D_{\text{Rx},m}^{\mathsf{x}} \hat{\phi}_{\ell}^{\mathsf{x}} + D_{\text{Rx},m}^{\mathsf{y}} \hat{\phi}_{\ell}^{\mathsf{y}} \right)} \cfaoa\!\left(\nu D_{\text{Rx},m}^{\mathsf{x}} \right) \, \cfaoa\!\left(\nu D_{\text{Rx},m}^{\mathsf{y}} \right) \,.
\end{align} 
\end{subequations}
}
Note that the estimated steering vector $\steeraoaest_\ell$ and the expected value of the true steering vector $\E_{\erraoa_\ell^{\mathsf{x}},\erraoa_\ell^{\mathsf{y}}} \left\{\steeraoa_\ell \right\}$ are not the same, i.e., $\steeraoaest_\ell \ne \E_{\erraoa_\ell^{\mathsf{x}},\erraoa_\ell^{\mathsf{y}}} \left\{\steeraoa_\ell \right\}$.

Similarly, the expected value of the $n$th element of $\steeraod_\ell$ can be determined as
\rew{
\begin{align}\label{eq:expec_b}
    \E_{\erraod_\ell^{\mathsf{x}},\erraod_\ell^{\mathsf{y}}} \left\{b_{\ell,n}\right\} =  e^{j\nu\left(D_{\text{Tx},n}^{\mathsf{x}} \hat{\Phi}_{\ell}^{\mathsf{x}} + D_{\text{Tx},n}^{\mathsf{y}} \hat{\Phi}_{\ell}^{\mathsf{y}} \right)} \cfaod\!\left(\nu D_{\text{Tx},m}^{\mathsf{x}} \right) \, \cfaod\!\left(\nu D_{\text{Tx},n}^{\mathsf{y}} \right)
\end{align}
where  $\cfaod(\nu D_{\text{Tx},m}^{\mathsf{x}} )$ and $\cfaod(\nu D_{\text{Tx},n}^{\mathsf{y}})$ are the \acp{cf} of $\erraod_\ell^{\mathsf{x}}$ and $\erraod_\ell^{\mathsf{y}}$ evaluated at $\nu D_{\text{Tx},m}^{\mathsf{x}}$ and $\nu D_{\text{Tx},n}^{\mathsf{y}}$, respectively.}

In the following, the expected value for $a_{\ell,m}$ in \eqref{eq:expec_a} is evaluated for the special cases of uniform and Gaussian distributed estimation errors. The expected value for $b_{\ell,n}$ can be obtained in the same manner and is thus omitted here.

\paragraph{Uniform Distribution}
Let \rew{$\erraoa_\ell^{\mathsf{x}}, \erraoa_\ell^{\mathsf{y}}\sim \mathcal{U}(-\erraoa_{\max}, \erraoa_{\max})$} be uniformly distributed in the interval $[-\erraoa_{\max}, \erraoa_{\max}]$. Then, $\E_{\erraoa_\ell^{\mathsf{x}},\erraoa_\ell^{\mathsf{y}}} \{a_{\ell,m}\}$ can be directly evaluated as \cite{Bazzi.Xu.2016,Mao.etal.2015}
\rew{
\begin{subequations}\label{eq:expec_a_uni}
   \begin{align}
    \E_{\erraoa_\ell^{\mathsf{x}},\erraoa_\ell^{\mathsf{y}}} \left\{a_{\ell,m} \right\}\big|_{\erraoa_\ell\sim \mathcal{U}} &= \int_{\hat{\phi}_\ell^\mathsf{y}-\erraoa_{\max}}^{\hat{\phi}_\ell^\mathsf{y}+\erraoa_{\max}} \int_{\hat{\phi}_\ell^\mathsf{x}-\erraoa_{\max}}^{\hat{\phi}_\ell^\mathsf{x}+\erraoa_{\max}} \frac{1}{\left(2\erraoa_{\max}\right)^2} e^{j\nu\left(D_{\text{Rx},m}^{\mathsf{x}} \phi_{\ell}^{\mathsf{x}} + D_{\text{Rx},m}^{\mathsf{y}} \phi_{\ell}^{\mathsf{y}} \right)} \,\mathrm{d} \phi_\ell^\mathsf{x} \mathrm{d}\phi_\ell^\mathsf{y} \\
    &= \int_{\hat{\phi}_\ell^\mathsf{x}-\erraoa_{\max}}^{\hat{\phi}_\ell^\mathsf{x}+\erraoa_{\max}}
    \frac{1}{2\erraoa_{\max}} e^{j\nu D_{\text{Rx},m}^{\mathsf{x}} \phi_{\ell}^{\mathsf{x}}} \,\mathrm{d} \phi_\ell^\mathsf{x} \cdot  \int_{\hat{\phi}_\ell^\mathsf{y}-\erraoa_{\max}}^{\hat{\phi}_\ell^\mathsf{y}+\erraoa_{\max}}
    \frac{1}{2\erraoa_{\max}} e^{j\nu D_{\text{Rx},m}^{\mathsf{y}} \phi_{\ell}^{\mathsf{y}}} \,\mathrm{d} \phi_\ell^\mathsf{y} \\
    &= e^{j\nu D_{\text{Rx},m}^{\mathsf{x}} \hat{\phi}_{\ell}^{\mathsf{x}}} 
    \sinc{\nu D_{\text{Rx},m}^{\mathsf{x}}\erraoa_{\max}}  \cdot
    e^{j\nu D_{\text{Rx},m}^{\mathsf{y}} \hat{\phi}_{\ell}^{\mathsf{y}}} 
    \sinc{\nu D_{\text{Rx},m}^{\mathsf{y}}\erraoa_{\max}}\\
    &= e^{j\nu\left(D_{\text{Rx},m}^{\mathsf{x}} \hat{\phi}_{\ell}^{\mathsf{x}} + D_{\text{Rx},m}^{\mathsf{y}} \hat{\phi}_{\ell}^{\mathsf{y}} \right)} \cfuni\!\left(\nu D_{\text{Rx},m}^{\mathsf{x}} \right)\, \cfuni\!\left(\nu D_{\text{Rx},m}^{\mathsf{y}} \right)\,.
\end{align} 
\end{subequations}
}
Thus, the \ac{cf} of a uniform distribution is the sinc-function, 
i.e., $\cfuni(t)=\sinc{t\erraoa_{\max}}=\sin(t\erraoa_{\max})/(t\erraoa_{\max})$.

\paragraph{Gaussian Distribution}
The \ac{cf} of a zero mean Gaussian distributed random variable with variance $\sigma^2$ is also a Gaussian function, i.e., $\cfgau(t)=e^{-\frac{t^2\sigma^2}{2}}$ \cite[Example~8.5]{Kobayashi.Mark.Turin.2011}.
Thus, if the estimation error has Gaussian distribution, i.e., \rew{$\erraoa_\ell^{\mathsf{x}}, \erraoa_\ell^{\mathsf{y}}\sim\mathcal{N}(0,\sigma_{\erraoa}^2)$}, \eqref{eq:expec_a} evaluates to
\rew{
\begin{subequations}\label{eq:expec_a_gau}
 \begin{align}
    \E_{\erraoa_\ell^{\mathsf{x}},\erraoa_\ell^{\mathsf{y}}} \left\{a_{\ell,m}\right\}\big|_{\erraoa_\ell\sim \mathcal{N}} &= e^{j\nu\left(D_{\text{Rx},m}^{\mathsf{x}} \hat{\phi}_{\ell}^{\mathsf{x}} + D_{\text{Rx},m}^{\mathsf{y}} \hat{\phi}_{\ell}^{\mathsf{y}} \right)} 
    \cfgau\!\left(\nu D_{\text{Rx},m}^{\mathsf{x}}\erraoa_{\max}\right)\,
    \cfgau\!\left(\nu D_{\text{Rx},m}^{\mathsf{y}}\erraoa_{\max}\right) \\
    &= e^{j\nu\left(D_{\text{Rx},m}^{\mathsf{x}} \hat{\phi}_{\ell}^{\mathsf{x}} + D_{\text{Rx},m}^{\mathsf{y}} \hat{\phi}_{\ell}^{\mathsf{y}} \right)}
    e^{-\frac{\left(\nu D_{\text{Rx},m}^{\mathsf{x}}\sigma_\erraoa\right)^2}{2}}
    e^{-\frac{\left(\nu D_{\text{Rx},m}^{\mathsf{y}}\sigma_\erraoa\right)^2}{2}} \,.
\end{align}
\end{subequations}
}
In the following, the \ac{cf} is used to derive the autocorrelation matrices of the steering vectors, which are necessary for optimum precoder and equalizer design.

\subsection{Precoder}\label{sec:robust_pc}
As in Section \ref{sec:ad_pc}, the robust precoder $\PC_{\text{rob}}$ is also designed to maximize the average received power, i.e.,
\begin{align}\label{eq:opt_pc_rob}
		\PC_{\text{rob}} \in \argmax_{\forall \ell: \tr{\PC_\ell^H\PC_\ell}\le \rho}\enskip \minrew{\E_{\left\{\alpha_{\ell}, \erraod_\ell^{\mathsf{x}}, \erraod_\ell^{\mathsf{y}}\right\}_{\ell=1}^{\NS}} \!\left\{\tr{\PC^H\chest^H\chest\PC}\right\}}
\end{align}
However, the expected value is now also taken w.r.t. the estimation errors. Thus, the objective function in \eqref{eq:opt_pc_rob} evaluates to
\rew{
\begin{subequations}\label{eq:objective_pc_rob}
	\begin{align}
		\E_{\left\{\alpha_{\ell}, \erraod_\ell^{\mathsf{x}}, \erraod_\ell^{\mathsf{y}}\right\}_{\ell=1}^{\NS}} \!\left\{\tr{\PC^H\chest^H\chest\PC}\right\} &= 
		\Nr\tr{\E_{\left\{ \erraod_\ell^{\mathsf{x}}, \erraod_\ell^{\mathsf{y}}\right\}_{\ell=1}^{\NS}} \!\left\{\PC^H\Steeraod\Sigma_{|\alpha|^2}\Steeraod^H\PC \right\}} \\
		&= \Nr\tr{\sum_{\ell=1}^{\NS} \sigma_{\alpha_\ell}^2 \E_{\erraod_\ell^{\mathsf{x}}, \erraod_\ell^{\mathsf{y}}}  \left\{\PC_\ell^H\steeraod_\ell\steeraod_\ell^H\PC_\ell\right\}} \\
		&= \Nr\sum_{\ell=1}^{\NS} \sigma_{\alpha_\ell}^2\tr{\PC_\ell^H\vekR_{\steeraod_\ell}\PC_\ell} \label{eq:objective_pc_rob_alt} 
	\end{align}
\end{subequations} 
}
where $\vekR_{\steeraod_\ell}=\E_{\erraod_\ell^{\mathsf{x}}, \erraod_\ell^{\mathsf{y}}} \left\{\steeraod_\ell\steeraod_\ell^H \right\}$ is the autocorrelation matrix of the steering vector $\steeraod_\ell$. 
With the \ac{cf} of the estimation error $\cfaod(t)$ from Section~\ref{sec:cf}, the $(m,n)$th element of the autocorrelation matrix $[\vekR_{\steeraod_\ell}]_{m,n}$ can be written as (see, e.g., \eqref{eq:expec_b})
\rew{
\begin{subequations}
\begin{align}\label{eq:corr_aod}
	\left[\vekR_{\steeraod_\ell}\right]_{m,n} &=  \E_{\erraod_\ell^{\mathsf{x}}, \erraod_\ell^{\mathsf{y}}} \!\left\{e^{-j \nu\left(D_{\text{Tx},m}^{\mathsf{x},\ell} \Phi_{\ell}^{\mathsf{x}} + D_{\text{Tx},m}^{\mathsf{y},\ell} \Phi_{\ell}^{\mathsf{y}} - D_{\text{Tx},n}^{\mathsf{x},\ell} \Phi_{\ell}^{\mathsf{x}} - D_{\text{Tx},n}^{\mathsf{y},\ell} \Phi_{\ell}^{\mathsf{y}} \right)} \right\} \\
	\begin{split}
	&= 
    e^{-j\nu\left(\left(D_{\text{Tx},m}^{\mathsf{x}}-D_{\text{Tx},n}^{\mathsf{x}}\right) \hat{\Phi}_{\ell}^{\mathsf{x}} + \left(D_{\text{Tx},m}^{\mathsf{y}}-D_{\text{Tx},n}^{\mathsf{y}}\right)  \hat{\Phi}_{\ell}^{\mathsf{y}} \right)} \\
    &\phantom{{}-{}} \cdot \cfaod\!\left(\nu \left(D_{\text{Tx},m}^{\mathsf{x}}-D_{\text{Tx},n}^{\mathsf{x}}\right) \right) \, \cfaod\!\left(\nu \left(D_{\text{Tx},m}^{\mathsf{y}}-D_{\text{Tx},n}^{\mathsf{y}}\right) \right)
	\end{split}
\end{align}
\end{subequations}
}
For the special cases of uniform or Gaussian distributed estimation error $\erraod_\ell$, the elements of $\vekR_{\steeraod_\ell}$  can be determined in the same way as in \eqref{eq:expec_a_uni} or \eqref{eq:expec_a_gau}.
Now, we can formulate the robust precoder, as in the following proposition.

%

\begin{proposition}\label{prop:pc_robust}
    Let $\pc_{\ell,\mathrm{rob}}$ be a scaled eigenvector corresponding to the largest eigenvalue of $\vekR_{\steeraod_\ell}$, such that $\pc_{\ell,\mathrm{rob}}^H\pc_{\ell,\mathrm{rob}}=\rho$, for each satellite $\ell$. Then an optimal precoder for \eqref{eq:opt_pc_rob} is obtained.
\end{proposition}
\begin{IEEEproof}[Proof sketch]
    The sum in \eqref{eq:objective_pc_rob_alt} is maximized if each term is maximized independently. Thus, the optimization problem \eqref{eq:opt_pc_rob} \rew{is equivalent to}
\begin{align}\label{eq:opt_pc_rob_alt}
        \rew{\forall \ell :} \min_{\PC_\ell}\enskip -\tr{\PC_\ell^H\vekR_{\steeraod_\ell}\PC_\ell} 
        \quad\text{s.t.}\quad \tr{\PC_\ell^H\PC_\ell} \le \rho .
\end{align}
Following the same steps as for Proposition \ref{prop:pc}, the optimum precoder $\PC_{\ell}'$ must satisfy
    \rew{$\vekR_{\steeraod_\ell}\PC_{\ell}' = \lagrange_{\ell}'\PC_{\ell}'$. 
Consider $\lagrange_{\ell}'$ to be the maximum eigenvalue of $\vekR_{\steeraod_\ell}$. Then, this is satisfied by the proposed precoder $\PC_{\ell,\text{rob}} = [\nullvec_{\Nt\times(\ell-1)}, \pc_{\ell,\text{rob}}, \nullvec_{\Nt\times(\NS-\ell)}]$ and the optimum of \eqref{eq:opt_pc_rob_alt} is achieved}, i.e.,
\begin{align}
	-\tr{\PC_{\ell,\text{rob}}^H\vekR_{\steeraod_\ell}\PC_{\ell,\text{rob}}} = -\lagrange_{\ell}'\pc_{\ell,\text{rob}}^H\pc_{\ell,\text{rob}} = -\lagrange_{\ell}'\rho \,.
\end{align}
\end{IEEEproof}
In Fig. \ref{fig:beampattern}, the radiation pattern is shown for three possible precoding approaches with imperfect \ac{csit}. \rew{For the sake of better visibility, only the error in $\mathsf{x}$-direction is considered and $\phi_\ell^\mathsf{y}=0$, and consequently, $\aod_\ell^{\text{az}}=0$, is perfectly known. Furthermore, the satellite $\ell$ is equipped with a \ac{ula} of $\Nt=60$ antennas, deployed along the local $\mathsf{x}$-axis.} 
The blue line shows the pattern if the power is radiated into the direction of the estimated \ac{aod} $\estaod^{\mathrm{el}} = 90^\circ$, i.e., $\pc_\ell=\steeraodest_\ell$. Following the terminology of \cite{book:Dietrich.2008}, this approach is called the heuristic approach because the estimation error is not considered at all and the estimated angle is assumed to be the true one.
The orange line shows the radiation pattern, if the precoding vector is chosen to be the expected value of the steering vector, i.e., $\pc_\ell=\E_{\erraod_\ell^{\mathsf{x}}}\{\steeraod_\ell\}$, and
the green line shows the radiation pattern for the optimum precoder, i.e., $\pc_\ell=\pc_{\ell,\text{rob}}$.
Fig. \ref{fig:beampattern_uni} shows the radiation pattern under the assumption of uniform phase errors with $\erraod_\ell^{\mathsf{x}}\sim\mathcal{U}(-0.05,0.05)$ and Fig. \ref{fig:beampattern_gau} for a Gaussian distributed error $\erraod_\ell^{\mathsf{x}}\sim\mathcal{N}(0,(2\cdot0.05)^2/12)$. Thus, the variance of the estimation error is in both cases the same.
It can be observed that the heuristic approach, i.e., the blue lines, leads to the most narrow radiation pattern with the highest peak gain. The second approach gives a much broader main lobe, at the cost of a lower peak gain. Furthermore, a lot of power is transmitted into regions unlikely to contain the \ac{rx}.
With the proposed robust precoder, which is represented by the green line, the main lobe is a little broader than for the heuristic approach. However, the side lobes are further reduced.
This becomes clearly visible for the case of uniform estimation error, i.e., the left picture. A phase error of $\erraod_{\max}=0.05$ at $\estaod^{\mathrm{el}}=90^\circ$ corresponds to an angular error $\estaod^{\mathrm{el}} - \aodl^{\mathrm{el}} \approx 2.866^\circ$. 
For the orange line, there is almost constant receive power at the \ac{rx} over the possible angular region. However, the true \ac{aod} must be $87.134^\circ \le \aodl^{\mathrm{el}} \le 92.866^\circ$. Thus, radiating outside this region is a waste of energy and may even  cause interference to other systems. Given the proposed robust precoder, only a small amount of the total power is radiated towards $\aodl^{\mathrm{el}} \le 87.134^\circ$ and $\aodl^{\mathrm{el}} \ge 92.866^\circ$. For the Gaussian case, the probability that the true angle is in the interval $[87.134^\circ,92.866^\circ]$ is approximately $\text{Pr}\left\{87.134^\circ \le \aodl^{\mathrm{el}} \le 92.866^\circ\right\}|_{\erraoa_\ell\sim\mathcal{N}} \approx 92\%$ and the sidelobes are less reduced compared to the uniform case.

\begin{figure}
	\centering
	\begin{subfigure}{0.45\columnwidth}
		\input{./figure/beampattern_uniform_centered.tikz}
			\vspace{-2ex}
		\caption{Uniform distributed error with $100\%$ confidence interval}
		\label{fig:beampattern_uni}
	\end{subfigure}
	\hfill
	\begin{subfigure}{0.45\columnwidth}
		\input{./figure/beampattern_gauss_centered.tikz}
			\vspace{-2ex}
		\caption{Gaussian distributed error with $95\%$ confidence interval}
		\label{fig:beampattern_gau}
	\end{subfigure}
	\caption{Radiation pattern for three different precoder $\pc_\ell\in\{\steeraodest_\ell,\E_{\erraod_\ell^{\mathsf{x}}}\{\steeraod_\ell\},\pc_{\ell,\text{rob}}\}$ and estimated \ac{aod} $\hat{\aod}_\ell=90^\circ$.}
	\label{fig:beampattern}%
\end{figure}

\subsection{Equalizer}\label{sec:robust_eq}
In this section, the linear equalizer that maximizes the mean \ac{sinr} with respect to the error model in \eqref{eq:est_error_angles} is derived.
Starting from \eqref{eq:ch_appr_sat} and under the assumption that the geometry based precoder with perfect \ac{aod} knowledge \eqref{eq:pc_ad} is used, the received signal in \eqref{eq:y_sum} becomes
\begin{align}\label{eq:y_appr}
    \veky \approx \sum_{\ell=1}^{\NS} \chest_\ell\pc_{\ell,\text{geo}} s_\ell + \vekn 
    = \sum_{\ell=1}^{\NS} \alpha_\ell\sqrt{\Nt\rho}\veka_\ell s_\ell + \vekn.
\end{align}    
%
Then, the optimization problem to determine the robust equalizer becomes
\begin{subequations}\label{eq:opt_eq_rob}
    \begin{align}
    \eq_{\ell,\text{rob}} &\in \argmax_{\eq_\ell}\, \E_{\left\{\alpha_{\ell'}, \erraoa_{\ell'}^{\mathsf{x}}, \erraoa_{\ell'}^{\mathsf{y}}\right\}_{\ell'=1}^{\NS}}\! \left\{\frac{\left\vert\eqsat^H\chest_{\ell}\pc_{\ell,\text{geo}}\right\vert^2} {\sum_{i\neq\ell} \left\vert \eqsat^H\chest_{i}\pc_{i,\text{geo}} \right\vert^2 + \nvar\eqsat^H\eqsat} \right\}\\
    	&= \argmax_{\eq_\ell}\frac{\sigma_{\alpha_\ell}^2 \eq_\ell^H \E_{\erraod_\ell^{\mathsf{x}}, \erraod_\ell^{\mathsf{y}}} \left\{\steeraoa_\ell\steeraoa_\ell^H\right\}\eq_\ell} {\eqsat^H\left(\sum_{i\neq\ell} \sigma_{\alpha_i}^2 \E_{\erraod_i^{\mathsf{x}}, \erraod_i^{\mathsf{y}}} \left\{\steeraoa_i\steeraoa_i^H \right\} + \naltvar\vekI_{\Nr}\right)\eqsat} \\
    	&= \argmax_{\eq_\ell}\frac{\sigma_{\alpha_\ell}^2 \eq_\ell^H\vekR_{\steeraoa_\ell}\eq_\ell} {\eq_\ell^H\left(\sum_{i\neq\ell} \sigma_{\alpha_i}^2\vekR_{\steeraoa_i} + \naltvar\vekI_{\Nr}\right)\eqsat}\,.
    \end{align}
\end{subequations}
where $\vekR_{\steeraoa_\ell} = \E_{\erraod_\ell^{\mathsf{x}}, \erraod_\ell^{\mathsf{y}}} \{\steeraoa_\ell\steeraoa_\ell^H\}$ is the autocorrelation of the steering vector $\steeraoa_\ell$. Similiar to \eqref{eq:corr_aod}, $\vekR_{\steeraoa_\ell}$ is also a Toeplitz matrix and its $(m,n)$th element is given by
\rew{
\begin{align}
\begin{split}
  	\left[\vekR_{\steeraoa_\ell}\right]_{m,n} &= e^{j\nu\left(\left(D_{\text{Rx},m}^{\mathsf{x}}-D_{\text{Rx},n}^{\mathsf{x}}\right) \hat{\phi}_{\ell}^{\mathsf{x}} + \left(D_{\text{Rx},m}^{\mathsf{y}}-D_{\text{Rx},n}^{\mathsf{y}}\right)  \hat{\phi}_{\ell}^{\mathsf{y}} \right)} \\
	&\phantom{{}-{}} \cdot \cfaoa\left(\nu \left(D_{\text{Rx},m}^{\mathsf{x}}-D_{\text{Rx},n}^{\mathsf{x}}\right) \right) \cfaoa\left(\nu \left(D_{\text{Rx},m}^{\mathsf{y}}-D_{\text{Rx},n}^{\mathsf{y}}\right) \right) .    
\end{split}
\end{align}
}%
The proposed equalizer is stated in the following proposition.

\begin{proposition}\label{prop:robust_eq}
    The receive vector for satellite $\ell$ \rew{that} maximizes the mean \ac{sinr} is proportional to the eigenvector corresponding to the largest eigenvalue of 
    $\left(\sum_{i\neq\ell} \sigma_{\alpha_i}^2 \vekR_{\steeraoa_i} + \naltvar\vekI_{\Nr}\right)^{-1}\corrmtx_{\steeraoa_\ell}$.
\end{proposition}

\begin{IEEEproof}[Proof sketch]
This proof follows along the lines of that of Proposition~\ref{prop:eq}. The objective function in \eqref{eq:opt_eq_rob} is again a generalized Rayleigh quotient. According to Lemma \ref{lem:rayleigh} in Appendix~\ref{sec:rayleigh}, it is maximized by the generalized eigenvector corresponding to the largest eigenvalue.
\end{IEEEproof}

\rew{In contrast to the case with perfectly known position, no closed form solution for the precoder and equalizer could be found. Instead, it requires the numerical computation of eigenvectors. This leads to slightly higher computational complexity compared to the previous case.}
Furthermore, it should be noted that only in case of perfect \ac{csir}, the achievable rate is maximized by maximizing the \ac{sinr} for each satellite \cite{Tse2005}.
However, for imperfect \ac{csir}, there is no analytic expression for the achievable rate known, which makes it difficult to \rew{optimize the equalizer w.r.t. the achievable rate} \cite{book:Dietrich.2008}.

\section{Optimal Inter-Satellite Distance}\label{sec:d_sat}
\begin{figure}
    \centering
	\includegraphics[height=2.3in]{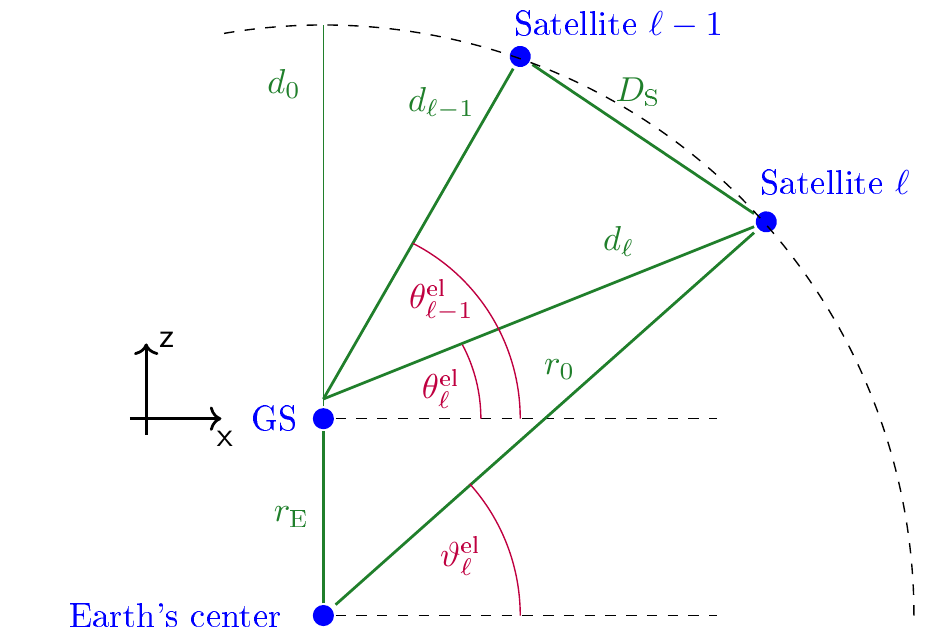}
	\caption{Geometric relation between two Satellites, $\ell$ and $\ell-1$, and the \ac{rx} \rew{for $\aoa_{\ell}^{\text{az}}=\aoa_{\ell-1}^{\text{az}}=0$}.}
	\label{fig:aoa}
\end{figure}

\rew{In this section, an analytical solution for the optimal inter-satellite distance is derived. To keep the analysis tractable, we make a few simplifying assumptions. In particular, we assume that the azimuth angles $\aoa_{\ell}^{\text{az}}$, for all satellites $\ell$, is zero and that the satellite swarm is flying in a trail formation with constant inter-satellite spacing $\DS$. This implies that the satellites follow a common orbit with altitude $d_0$ and the \ac{rx}'s antenna array is effectively seen as a linear array in the orbital plane. We further assume that the \ac{rx}'s antennas are uniformly spaced with distance $\DARx$. Hence, without loss of generality (given these assumptions), we model the \ac{rx}'s antenna array as a \ac{ula} with $\Nr^\mathsf{x}$ antennas, and the $\mathsf{xz}$-plane of the \ac{rx}-centered coordinate frame is chosen such that it is aligned with the orbital plane.}
\rew{Besides the \ac{rx}-centered coordinate frame, we also require an \ac{eci} coordinate frame whose $\mathsf{xz}$-plane is aligned with the orbital plane. With the Earth's radius being $\RE=\SI{6371}{\kilo\meter}$, the orbital radius is $r_0=\RE + d_0$ and the position of satellite $\ell$ in the \ac{eci} coordinate frame is given by the triplet $(r_0, \vartheta_\ell^{\text{el}}, 0)$. Consequently, the \ac{rx} is located at position $(\RE, \pi/2, 0)$. This setup is illustrated in Fig.~\ref{fig:aoa}.}

Considering the triangle between satellite $\ell$, the \ac{rx} and the Earth's center as illustrated in Fig.~\ref{fig:aoa}, we obtain from the law of sines 
\minrew{\begin{align}
	\vartheta_\ell^{\text{el}} &= \aoal^{\text{el}} + \arcsin\left(\frac{\RE}{r_0} \cos(\aoal^{\text{el}})\right) \label{eq:vartheta} \\
	d_\ell &= r_0\frac{\cos\left(\vartheta_\ell^{\text{el}}\right)}{\cos\left(\aoal^{\text{el}}\right)}\label{eq:d_sin}\,. 
\end{align}}
%
Then, the inter-satellite distance $\DS$ is obtained from the law of cosines as
\minrew{
\begin{align}\label{eq:D_S}
\DS = \sqrt{d_{\ell}^2 + d_{\ell-1}^2 - 2d_{\ell}d_{\ell-1}\cos\left(\aoa_{\ell-1}^{\text{el}} - \aoal^{\text{el}}\right)}
\end{align}}
where \minrew{$\vartheta_{\ell-1}^{\text{el}}$} and $d_{\ell-1}$ are given analogously to \eqref{eq:vartheta} and \eqref{eq:d_sin}, respectively.


\rew{Now, the relation between the inter-satellite distance $\DS$ and the achievable rate is analyzed based on the channel approximation \eqref{eq:ch_appr_sat}. We further approximate the channel gains of all satellites to be equal, i.e.,  $\sigma_{\alpha_\ell}^2 \approx \sigma_{\alpha}^2$, for all $\ell$. This approximation is valid for small inter-satellite distances but might be not entirely accurate for very large distances.}
\rew{Assuming the precoder $\PC_{\text{geo}}$ from \eqref{eq:pc_ad} is employed at the satellites and we have perfect \ac{csir},} the ergodic rate $\tilde{R}$ for $\chest$ can be upper bounded as
	\begin{align}\label{eq:rate_approx}
		\tilde{R} &\le \minrew{\log_2\left\vert \vekI_{\Nr} + \frac{1}{\nvar} \E_{\left\{\alpha_\ell\right\}_{\ell=1}^{\NS}} \left\{ \chest\PC_\mathrm{geo}\PC_\mathrm{geo}^H\chest^H \right\} \right\vert} 
		= \minrew{\log_2\left\vert \vekI_{\Nr} + \frac{\sigma_{\alpha}^2}{\naltvar}\vekA\vekA^H \right\vert}\,. 
	\end{align}
%
Thus, the achievable rate over the channel $\chest$ depends on the matrix $\vekA$, which is composed of the steering vectors $\{\veka_\ell\}_{\ell=1}^{\NS}$.

Due to the trail formation, the swarm is fully described by two parameters: The inter-satellite distance $\DS$ and the number of satellites $\NS$.
Choosing a proper inter-satellite distance $\DS$ is crucial, as it directly impacts the angular spread of the elevation angles between the satellites,  which can be used to tune the matrix $\vekA$ such that the achievable rate is maximized. Therefore, we aim to find an inter-satellite distance $D_{\text{S,opt}}$ such that
\begin{align}\label{eq:opt_ds}
    D_{\text{S,opt}} &\in \argmax_{\DS}\enskip \minrew{\log_2\left\vert \vekI_{\Nr} + \frac{\sigma_{\alpha}^2}{\naltvar}\vekA\vekA^H \right\vert} \,.
\end{align}
Solving this optimization problem is difficult due to the lack of a clear functional relationship between $\vekA$ and $\DS$. A sufficient optimality condition that partially characterizes the solution space of \eqref{eq:opt_ds} is stated next.

\begin{proposition}\label{prop:orth_channels}
    Let $k$ be a positive integer such that it is no multiple of $\Nr^\mathsf{x}$.
    If the elevation angles satisfy
	\begin{align}\label{eq:orth_cond}
		\forall \ell : \forall i \neq \ell: \minrew{|\cos(\aoa_\ell^\mathrm{el}) - \cos(\aoa_{i}^\mathrm{el})| = \frac{2\pi k}{\nu\DARx\Nr^\mathsf{x}}},
	\end{align}
	then \eqref{eq:rate_approx} is maximized.
\end{proposition}

\begin{IEEEproof}
	Observe that \eqref{eq:rate_approx} is equivalent to
	\begin{equation}\label{eq:rate_approx_ext}
		\tilde{R} \le \minrew{\log_2\left\vert \vekI_{\NS} + \frac{\sigma_{\alpha}^2}{\naltvar}\vekA^H \vekA \right\vert
		= \log_2\left(\prod_{\ell=1}^{\NS}\left( 1 + \frac{\tilde{\lambda_\ell}\sigma_{\alpha}^2}{\naltvar} \right)\right)}
	\end{equation}
	where $\tilde{\lambda}_\ell$ are the positive eigenvalues of $\vekA^H \vekA$ \cite{Telatar.1999}. Keeping the trace of $\vekA^H \vekA$ constant, this is maximized if all eigenvalues have the same value \cite[Thm.~2.21]{Jorswieck2007}. In other words, any $N_S\times N_S$ matrix $\vekZ = \vekA^H \vekA$ maximizing \eqref{eq:rate_approx_ext} has a single eigenvalue $\tilde{\lambda}$ with multiplicity $N_S$.
	
	Further, observe that $\vekZ$ is a normal matrix. By \cite[Thm.~2.5.4]{Horn1990}, $\vekZ$ is similar to a diagonal matrix, i.e., there exists a nonsingular matrix $\vekS$ such that $\vekS^{-1} \vek{\Lambda} \vekS = \vekZ$ with $\vek{\Lambda}$ diagonal. Since similar matrices have the same eigenvalues \cite[Cor.~1.3.4]{Horn1990}, $\vek{\Lambda}$ must be $\tilde{\lambda} \vekI$. Then, for every nonsingular $\vekS$, we have $\vekZ = \vekS^{-1} \tilde{\lambda} \vekI \vekS = \tilde{\lambda} \vekS^{-1} \vekS = \tilde{\lambda} \vekI$. It follows that $\vekZ = \tilde{\lambda}\vekI$ is the unique maximizer of \eqref{eq:rate_approx_ext}.	
	For $\vekA^H\vekA$ to become a scaled identity matrix, its columns must satisfy $\veka_i^H\veka_i = \tilde\lambda$ and $\veka_i^H\veka_\ell = 0$ for all $i$ and $\ell \neq i$. From the definition of $\veka$, we obtain
	\begin{equation}
	    \veka_i^H\veka_\ell = \sum_{m=0}^{\Nr-1} e^{j\nu\DARx m\left(\minrew{ \cos(\aoa_\ell^\mathrm{el}) - \cos(\aoa_{i}^\mathrm{el}) }\right)}.
	\end{equation}
	This equals zero if $\minrew{\nu\DARx m\left( \cos(\aoa_\ell^\mathrm{el}) - \cos(\aoa_{i}^\mathrm{el}) \right)= 2\pi \frac{km}{\Nr^\mathsf{x}}}$ for some integer $k$ such that $k \neq n \minrew{\Nr^\mathsf{x}}$ for all integer $n$ \cite{Lin.Gao.Jin.Li.2017}. This is equivalent to the condition above.
	Finally, observe that
	\begin{equation}
	    \veka_i^H\veka_i = \sum_{m=0}^{\Nr-1} e^{j\nu\DARx m\left(\minrew{ \cos(\aoa_i^\mathrm{el}) - \cos(\aoa_{i}^\mathrm{el})} \right)} = \sum_{m=0}^{\Nr-1} e^{j0} = \minrew{\Nr^\mathsf{x}}.
	\end{equation}
\end{IEEEproof}

Consider now two neighbouring satellites $\ell$ and $\ell-1$.
According to Proposition \ref{prop:orth_channels}, the first maximum is achieved if
\minrew{
\begin{subequations}\label{eq:DeltaPhi_gen}
	\begin{align}
		\frac{2\pi}{\nu\DARx\Nr^\mathsf{x}} &=\left|\cos\left(\theta_{\ell}^\mathrm{el}\right) - \cos\left(\theta_{\ell-1}^\mathrm{el}\right) \right|    \Leftrightarrow\\
		 \theta_{\ell-1} &= \begin{cases}
		 	\arccos\left(\cos\left(\theta_{\ell}^\mathrm{el}\right) - \frac{2\pi}{\nu\DARx\Nr^\mathsf{x}}\right)	&\text{ for } \cos(\theta_{\ell}^\mathrm{el})\ge\cos(\theta_{\ell-1}^\mathrm{el})\\
		 	\arccos\left(\frac{2\pi}{\nu\DARx\Nr^\mathsf{x}} - \cos\left(\theta_{\ell}^\mathrm{el}\right) \right)	&\text{ for } \cos(\theta_{\ell-1}^\mathrm{el})\ge\cos(\theta_{\ell}^\mathrm{el})
		\end{cases}\,.\label{eq:theta_l-1}
	\end{align}
\end{subequations}}
Then,
plugging \eqref{eq:theta_l-1} in \eqref{eq:vartheta} and then in \eqref{eq:d_sin} and \eqref{eq:D_S}, respectively, one gets a solution for the optimal inter-satellite distance $D_{\text{S,opt}}$, which depends on the altitude $d_0$, the \minrew{elevation angle $\aoal^\mathrm{el}$} of satellite $\ell$, the number of receive antennas $\minrew{\Nr^\mathsf{x}}$ and the antenna spacing $\DARx$ at the \ac{rx}, as well as the wavenumber $\nu$. As there is little intuition to be gained from the explicit formula  for $D_{\text{S,opt}}$ \rew{\eqref{eq:D_S_opt}}, it is delegated to Appendix~\ref{sec:D_S}.

\begin{figure}
	\begin{subfigure}{0.47\columnwidth}
\begin{tikzpicture}

\definecolor{color0}{rgb}{0.12156862745098,0.466666666666667,0.705882352941177}
\definecolor{color1}{rgb}{1,0.498039215686275,0.0549019607843137}
\definecolor{color2}{rgb}{0.172549019607843,0.627450980392157,0.172549019607843}

\begin{axis}[
width=0.85\columnwidth,
height=2in,
legend cell align={left},
legend style={fill opacity=0.8, draw opacity=1, text opacity=1, draw=white!80!black},
log basis y={10},
scale only axis,
tick pos=left,
xlabel={Elevation angle $\theta_\ell^\mathrm{el}\;[^\circ]$},
xmajorgrids,
xmin=1, xmax=90,
xtick style={color=black},
ylabel={$D_{\text{S,opt}}\;[\text{km}]$},
ymajorgrids,
ymin=9.25435572039344, ymax=2000,
ymode=log,
ytick style={color=black}
]
\addplot [thick, color0, dashed]
table {%
1 1956.3083092464
2 1844.12299336245
3 1737.34144452408
4 1635.91679278339
5 1539.7735307661
6 1448.80983081797
7 1362.90029301165
8 1281.89900011394
9 1205.64275228461
10 1133.95435962805
11 1066.64588307959
12 1003.52173135725
13 944.38154162855
14 889.022792102289
15 837.243114316049
16 788.842290284945
17 743.623934217483
18 701.396869918975
19 661.976223345405
20 625.184255316794
21 590.850962554172
22 558.814476425367
23 528.921288526029
24 501.026330898378
25 474.992936657436
26 450.692704342747
27 428.005286668975
28 406.818121677951
29 387.026121715361
30 368.531333244156
31 351.242578309075
32 335.075086504171
33 319.950124572757
34 305.794629279368
35 292.540847922158
36 280.125989782568
37 268.491890916351
38 257.584693955082
39 247.354543988881
40 237.755301120585
41 228.744269900525
42 220.281945554238
43 212.331776688233
44 204.859943989469
45 197.835154311981
46 191.228449459988
47 185.013028923352
48 179.164085792203
49 173.658655067305
50 168.475473587304
51 163.594850809695
52 158.998549705952
53 154.669677061086
54 150.59258250126
55 146.752765609358
56 143.136790525559
57 139.732207467822
58 136.527480644647
59 133.511922069146
60 130.67563081892
61 128.009437320477
62 125.504852269311
63 123.154019827583
64 120.94967477063
65 118.88510328069
66 116.954107112083
67 115.150970876124
68 113.470432216535
69 111.907654667167
70 110.45820300362
71 109.11802091861
72 107.883410868324
73 106.751015953182
74 105.717803711641
75 104.781051720209
76 103.938334906545
77 103.187514495626
78 102.526728521678
79 101.954383850798
80 101.469149671291
81 101.06995242061
82 100.755972129735
83 100.526640177915
84 100.38163846301
85 100.320900005519
86 100.344611017844
87 100.453214484546
88 100.647415314742
89 100.92818714437
90 101.296780884262
};
\addlegendentry{$N_\text{r}^\mathsf{x}=10$}

\addplot [thick, color1, dashdotted]
table {%
1 1493.55664491765
2 1383.12687276432
3 1279.58226431011
4 1182.8065290867
5 1092.63174478677
6 1008.84475986823
7 931.194548810721
8 859.400056441001
9 793.158094014758
10 732.150912211813
11 676.053160589652
12 624.538035156921
13 577.282503865667
14 533.971575948407
15 494.301640920971
16 457.982945716076
17 424.741305058076
18 394.319153398457
19 366.476049580235
20 340.988740918957
21 317.650884233584
22 296.272509667286
23 276.679300496899
24 258.711749643295
25 242.22424196131
26 227.084100995113
27 213.170629899102
28 200.374168659405
29 188.595183527707
30 177.743399565519
31 167.736983237486
32 158.5017789279
33 149.970600933431
34 142.082580769185
35 134.782568395181
36 128.020585122092
37 121.75132540431
38 115.93370440382
39 110.530448053589
40 105.507722318827
41 100.834798413182
42 96.4837508468792
43 92.4291853439288
44 88.6479938497878
45 85.1191340477371
46 81.8234310024429
47 78.7433987471884
48 75.8630798226919
49 73.1679009569961
50 70.6445432468518
51 68.2808253596241
52 66.0655984209783
53 63.9886513879399
54 62.0406258289688
55 60.2129391438863
56 58.4977153568523
57 56.8877227062522
58 55.3763173368728
59 53.9573924731092
60 52.6253325175817
61 51.3749715785458
62 50.2015559820989
63 49.1007103723561
64 48.0684070449796
65 47.1009381970728
66 46.1948908100719
67 45.3471239125059
68 44.5547479962277
69 43.8151063839188
70 43.1257583671503
71 42.4844639535025
72 41.8891700786765
73 41.3379981549878
74 40.8292328414628
75 40.3613119335947
76 39.9328172819164
77 39.5424666589116
78 39.1891065031401
79 38.8717054778565
80 38.5893487892492
81 38.3412332165018
82 38.1266628125321
83 37.9450452403913
84 37.7958887160694
85 37.6787995339342
86 37.5934801562198
87 37.5397278530685
88 37.5174338844814
89 37.5265832205108
90 37.567254800789
};
\addlegendentry{$N_\text{r}^\mathsf{x}=30$}

\addplot [thick, color2]
table {%
1 971.935544819952
2 867.119638091372
3 772.912283424006
4 688.765382299025
5 613.979423269051
6 547.762647484643
7 489.285515298505
8 437.724672126873
9 392.294368410963
10 352.266060762864
11 316.978392152274
12 285.840160480629
13 258.328659855341
14 233.985283897641
15 212.409748193454
16 193.253832011188
17 176.215192073312
18 161.031557936356
19 147.475458761462
20 135.349532245248
21 124.48240889825
22 114.725134126882
23 105.948076475613
24 98.0382661173112
25 90.8971088906528
26 84.438425188703
27 78.5867682048119
28 73.275981527887
29 68.4479613615777
30 64.0515934811503
31 60.0418393479746
32 56.3789495615645
33 53.0277860745402
34 49.9572373739807
35 47.1397132010069
36 44.5507073911816
37 42.1684191240592
38 39.9734243151797
39 37.9483901074017
40 36.0778264554749
41 34.3478696761555
42 32.7460935818051
43 31.2613444474332
44 29.8835965990374
45 28.6038258678663
46 27.4138985444069
47 26.3064737976552
48 25.2749178079696
49 24.3132281037085
50 23.4159667984985
51 22.5782016029088
52 21.7954536360437
53 21.0636511929561
54 20.3790887353781
55 19.7383904695562
56 19.1384779582894
57 18.5765412852291
58 18.0500133514714
59 17.5565469380447
60 17.0939942137575
61 16.6603884081035
62 16.253927403762
63 15.8729590331213
64 15.5159678899489
65 15.1815634897953
66 14.8684696330184
67 14.5755148415029
68 14.3016237556655
69 14.0458093913429
70 13.8071661683247
71 13.584863632104
72 13.3781407998471
73 13.1863010694733
74 13.008707637629
75 12.8447793787229
76 12.6939871427562
77 12.5558504343915
78 12.4299344403443
79 12.315847375749
80 12.2132381240669
81 12.1217941478252
82 12.0412396507101
83 11.9713339737502
84 11.9118702109849
85 11.8626740320765
86 11.8236027011775
87 11.7945442835544
88 11.7754170328483
89 11.7661689539095
90 11.7667775374496
};
\addlegendentry{$N_\text{r}^\mathsf{x}=100$}
\end{axis}

\end{tikzpicture}
	\end{subfigure}
	\hfill
	\begin{subfigure}{0.47\columnwidth}
		\input{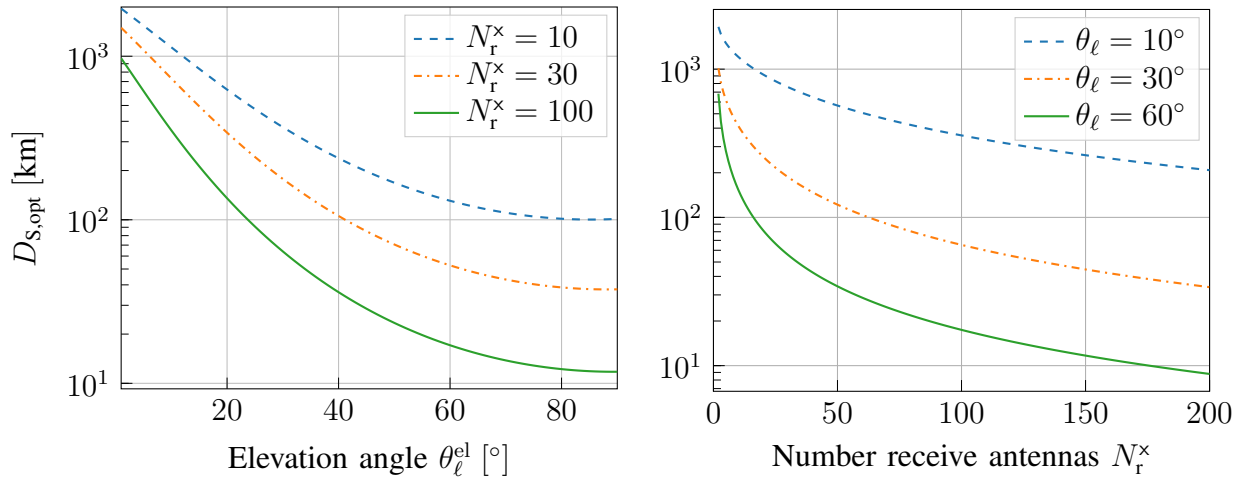}
	\end{subfigure}	
	\vspace{-4ex}
	\caption{Optimal inter-satellite distances $D_\text{S,opt}$ for different elevation angles $\theta_\ell^\mathrm{el}$ and for different number of receive antennas $\Nr^\mathsf{x}$ at altitude $d_0=600\,$km.}
	\label{fig:DvsTheta}%
\end{figure}

In Fig. \ref{fig:DvsTheta}, the dependency between the optimum inter-satellite distance $D_\text{S,opt}$, the \minrew{elevation angle $\aoal^\mathrm{el}$} and number of receive antennas \minrew{$\Nr^\mathsf{x}$} is shown for an altitude $d_0=\SI{600}{\kilo\meter}$ and $\nu\DARx=\pi$.
Note that, according to \eqref{eq:orth_cond}, the product $\minrew{\nu\DARx\Nr^\mathsf{x}}$ determines the spatial resolution. Therefore, increasing the antenna spacing $\DARx$ by a certain factor and decreasing the number of antennas $\minrew{\Nr^\mathsf{x}}$ by the same factor, gives the same results. However, a smaller number of antennas leads to a smaller array gain and if the antenna spacing $\DARx$ and the inter-satellite distance $\DS$ are both very large, spatial aliasing must be considered.

It can be seen in Fig. \ref{fig:DvsTheta} that the optimum inter-satellite distance $D_\text{S,opt}$ increases strongly as the elevation angle \minrew{$\theta_\ell^\mathrm{el}$} decreases.
\minrew{However, the satellites have distinct elevation angles, i.e., $\theta_\ell^\mathrm{el} \neq \theta_i^\mathrm{el}$ for all $i\neq \ell$. This means that the optimal distance between satellites $i$ and $i-1$ is different to the optimal distance between satellite $\ell$ and $\ell-1$.
Furthermore, the elevation angles changes over time. Thus, it is not possible to ensure orthogonal steering vectors between all satellites, i.e., $\steeraoa_i^H\steeraoa_\ell=0$ for all $i\ne\ell$, during the whole flight with a constant inter-satellite distance $\DS$.}
Adjusting the inter-satellite distance during the flight, however, requires additional fuel and increased complexity for flight control and should thus be avoided.
Nevertheless, as evaluated numerically in Section \ref{sec:simulation}, the channel capacity is not decreasing much after the first local optimum 
and, thus, a close-to-optimal heuristic is obtained by relaxing condition \eqref{eq:orth_cond}. In particular, the average rate over the whole flight is close to maximum if
\minrew{\begin{align}\label{eq:orth_cond_relaxed}
	\min_\ell \left|\cos(\aoa_\ell^\mathrm{el}) - \cos(\aoa_{\ell-1}^\mathrm{el})\right| \geq \frac{2\pi}{\nu\DARx\Nr^\mathsf{x}}
\end{align}}
holds during the transmission. This can be used as a guidance to find a proper inter-satellite distance $\DS$, if the minimum elevation angle is known a~priori.

Furthermore, the following relation between the optimal inter-satellite distance and the optimal precoder can be observed.
\begin{proposition}
    The proposed precoder in \eqref{eq:pc_ad} is capacity achieving for the channel $\chest$
	if \minrew{$|\cos(\aoa_\ell^\mathrm{el}) - \cos(\aoa_{i}^\mathrm{el})| = \frac{2\pi k}{\nu\DARx\Nr^\mathsf{x}}$} for all $\ell \neq i$ and $|\alpha_i|=|\alpha_\ell|=|\alpha|$.
\end{proposition}

\begin{IEEEproof}
	First, recall from the proof of Proposition~\ref{prop:pc} that the channel $\chest$ can be decomposed as
	$\chest = \Steeraoa\vekSig_\alpha\Steeraod^H$.
	Second, the capacity achieving precoder is given by the scaled dominant right singular vectors of the channel matrix \cite{Telatar.1999}.
	The \ac{svd} of the channel $\chest$ is defined as $\chest=\vektU\vektSig\vektV^H,$
	where $\vektU$ and $\vektV$ are unitary matrices and $\vektSig$ is a diagonal matrix with non-negative numbers on its main diagonal.
	
	The complex-valued diagonal matrix $\vekSig_\alpha$ is equivalent to
	   	\begin{align}
	        \vekSig_\alpha = \diag{|\alpha_1|,...,|\alpha_{\NS}|} \cdot \diag{e^{-j\phi_1},...,e^{-j\phi_{\NS}}} 
	        = \vekSig_{|\alpha|}\vekSig_{\phi} = \vekSig_{\phi}\vekSig_{|\alpha|} \,.
	    \end{align} 
    Furthermore, due to the block diagonal structure of $\Steeraod$, we have
    $\Steeraod^H\Steeraod = \Nt\vekI_{\NS}$
	and, if \eqref{eq:orth_cond} holds, we also have
	    $\Steeraoa^H\Steeraoa = \Nr^\mathsf{x}\vekI_{\NS}.$
	Thus, we can define the unitary matrices $\vektU$ and $\vektV$ as
	\begin{subequations}
	\begin{align}
	    \vektU &= \frac{1}{\sqrt{\Nr^\mathsf{x}}}\left[\Steeraoa,\vekt{u}_{\NS+1},...,\vekt{u}_{\Nr^\mathsf{x}}\right] 
	    \begin{bmatrix}
	        \vekSig_{\phi} & \nullvec_{(\Nr^\mathsf{x} - \NS) \times (\Nr^\mathsf{x} - \NS)} \\
	        \nullvec_{(\Nr^\mathsf{x} - \NS) \times (\Nr^\mathsf{x} - \NS)} & \vekI_{\Nr^\mathsf{x}-\NS}
	    \end{bmatrix}\\
	    \vektV &= \frac{1}{\sqrt{\Nt}}
	    \left[\Steeraod,\vekt{v}_{\NS+1},...,\vekt{v}_{\NTx}\right] 
	\end{align}	    
	\end{subequations}
    where $[\vekt{u}_{\NS+1},...,\vekt{u}_{\Nr^\mathsf{x}}]$ and $[\vekt{v}_{\NS+1},...,\vekt{v}_{\NTx}]$ are the left and right singular vectors belonging to the nullspace of $\chest$, respectively.
	Consequently, the singular values are given by $|\alpha_i|$, i.e.,
	\begin{align}
	    \vektSig = \sqrt{\Nr^\mathsf{x}\Nt}\begin{bmatrix}
	        \vekSig_{|\alpha|} & \nullvec_{\NS \times (\NTx-\NS)} \\
	        \nullvec_{(\Nr^\mathsf{x}-\NS)\times \NS} & \nullvec_{(\Nr^\mathsf{x}-\NS)\times (\NTx-\NS)}
	    \end{bmatrix} \,.
	\end{align}
	Consequently, the channel $\chest$ can be decomposed as
	    \begin{align}
	        \chest &= \vektU\vektSig\vektV^H 
	        = \Steeraoa \vekSig_{\phi}\vekSig_{|\alpha|} \Steeraod 
	        = \sqrt{\frac{\Nt}{\rho}}\Steeraoa \vekSig_{\phi}\vekSig_{|\alpha|}\PC_{\text{geo}}\,.
	    \end{align}
	Thus, the right singular vectors of $\chest$ are given by the proposed precoder \eqref{eq:pc_ad}. Furthermore, if $|\alpha_i|=|\alpha_\ell|=|\alpha|$ for all $i$ and $\ell$ holds, i.e., $\vekSig_{|\alpha|}=|\alpha|\vekI_{\NS}$, then all precoding vectors $\pc_{\ell,\text{geo}}$ must have the same power in order to achieve the capacity. This is given by the proposed precoder, as $\Vert\pc_{\ell,\text{geo}}\Vert_2^2=\Vert\pc_{i,\text{geo}}\Vert_2^2=\rho$ for all $i$ and $\ell$. 
\end{IEEEproof}

\rew{Note that several simplyfing assumptions have been made in this section. Therefore, the precoder and the inter-satellite distance are only optimal for distinct setups. In general they are sub-optimal. However, as shown in Section \ref{sec:simulation}, the numerical simulations show an almost optimal performance under more practical assumptions of the proposed precoder \eqref{eq:pc_ad} in combination with the linear equalizer \eqref{eq:eq_geo} at the optimal inter-satellite distance\eqref{eq:D_S_opt}, as well.
Furthermore,} compared to the optimal \ac{svd}-based precoder, \rew{the requirements on \ac{csi} acquisition are considerably relaxed and there is no need for any inter-satellite communication in order to determine the precoder matrix.}
Moreover, the computational complexity is also significantly reduced.

\section{Numerical Evaluations}\label{sec:simulation}
\subsection{Channel Model}\label{sec:channel}

We use the channel model recommened by
3GPP and ITU-R in \cite{3GPP.TR.38.811,ITUR.P.676,ITUR.P.835,ITUR.P.618,ITUR.P.531}. It is further assumed, that the \ac{rx} is placed such that the communication link is not blocked by other objects.
The carrier frequency is $\fc = \SI{20}{\giga\hertz}$ and \rew{we assume a pure \ac{los} connection between the \ac{rx} and each satellite. With the possibility of \ac{nlos} propagation, swarms with more satellites get a further advantage due to a higher probability that at least one satellite has a \ac{los} connection to the \ac{rx}. For the sake of fairer comparison, \ac{nlos} propagation is not considered.}
The $(m,n)$th element of the local channel matrix $\vekH_\ell$ is modeled as
\begin{align}\label{eq:channel_true}
	h_{m,n}^\ell &=\frac{1}{\sqrt{L_{m,n}^\ell}}e^{-j\left(\nu d_{m,n}^\ell + \phi_{\text{atm},\ell}\right)}
\end{align}
where $\phi_{\text{atm},\ell}\in[0,2\pi]$ is a uniformly distributed phase shift caused by the atmosphere, and $L$ is the path loss, which is given in decibel as \cite{3GPP.TR.38.811}
\begin{align}
		L_{m,n|\dB}^\ell  =\, 20\log_{10}\left(2\nu d_{m,n}\right) -\left(\GTx + \GRx \right)
		+ L_{\text{sf},\ell} + L_{\text{cl},\ell} + L_{\text{gas},\ell} + L_{\text{ts},\ell}   
\end{align}
where $L_{\text{sf},\ell}\sim\mathcal{N}(0,\sigma_{\text{sf},\ell}^2)$ and  $L_{\text{cl},\ell}$ are the shadow fading and clutter loss, repsectively. In LOS $L_{\text{cl},\ell}=\SI{0}{\dB}$ and $\sigma_{\text{sf},\ell}^2$ depends on the \ac{aoa}. The specific values can be found in \cite{3GPP.TR.38.811}, while in this paper the rural scenario has been considered.
$L_{\text{gas},\ell}$ includes atmospheric gas absorption described in \cite{ITUR.P.676} using the reference standard atmosphere \cite{ITUR.P.835}.
Eventually, $L_{\text{ts},\ell}$ includes the losses due to tropospheric scintillation, as summarized in \cite{ITUR.P.618,ITUR.P.531},
and $\GTx$ and $\GRx$ are the transmit and receive antenna gains, respectively.
\rew{In all simulations,  the transmission is assumed to start if the average elevation angle over all satellites $\aoa_{\text{mean}}^{\text{el}}=1/\NS\sum_{\ell=1}^{\NS}\aoal^{\text{el}} = 30^\circ$, i.e, the evaluation is done for $30^\circ\leq\aoa_{\text{mean}}^{\text{el}}\leq 150^\circ$. Furthermore, the noise power and altitude are set to $P_{\text{N,dB}}=\SI{-120}{\dBW}$ and 
$d_0=\SI{600}{\kilo\meter}$, respectively.}

\subsection{Inter-Satellite Distance}
In Section \ref{sec:d_sat}, the optimal inter-satellite distance $D_{\text{S,opt}}$ \rew{for a simplified scenario has been derived. In order to verify that derivation, we assume, in this subsection, $\aoa_\ell^{\text{az}}=0$ for all satellites $\ell$ and a constant inter-satellite distance $\DS$. 
As in Section \ref{sec:d_sat}, we model the \ac{rx} antenna array as an effective \ac{ula}. The number of receive antennas is $\Nr^{\mathsf{x}}=100$ and the receive antenna gain is $\GRx=\SI{20}{\dBi}$. In practical systems, this antenna gain can be achieved, e.g., with an $100\times 100$ \ac{ura}.
Likewise, we assume a \ac{ula} with $\Nt^{\mathsf{x}}=60/\NS$ antennas and an antenna gain of $\GTx=\SI{17.8}{\dBi}$ at each satellite. 
The sum transmit power of the satellite swarm is $\Ptx=\NS\rho=\SI{10}{\watt}$.
The normalization of the number of transmit antennas and the transmit power per satellite w.r.t. to the number of satellites $\NS$ is included to allow a fairer comparison between satellite swarms with different $\NS$.}
The channel capacity $R_{\text{opt}}$ from \eqref{eq:rate_opt} is considered as a benchmark. It requires \ac{svd} precoding and \ac{sic} at the \ac{rx} \cite{Telatar.1999,Tse2005}. 
This is compared against the geometric precoder \eqref{eq:pc_ad} with linear equalizer \eqref{eq:eq_geo}. The corresponding achievable rate is given by $R_{\text{lin}}$ in \eqref{eq:rate_sum}.

\begin{figure}[t]%
	\centering
	\begin{subfigure}[b]{0.47\columnwidth}
%
%
\definecolor{mycolor1}{rgb}{0.92941,0.69412,0.12549}%
\definecolor{mycolor2}{rgb}{0.85000,0.32500,0.09800}%
\definecolor{mycolor3}{rgb}{0.00000,0.44706,0.74118}%
\usetikzlibrary{spy}
\begin{tikzpicture}

\begin{axis}[%
width=0.85\columnwidth,
height=2in,
scale only axis,
xmin=0,
xmax=40,
xlabel={Inter-satellite  distance $D_\text{S}\,$[km]},
ymin=10,
ymax=17,
ylabel={Achievable rate [bps/Hz]},
axis background/.style={fill=white},
xmajorgrids,
xtick style={color=black},
ymajorgrids,
ytick style={color=black},
legend style={at={(0.98,0.01)}, anchor=south east, legend cell align=left, align=left, fill opacity=0.8, draw opacity=1, text opacity=1, draw=white!80!black}
]
\addplot [color=mycolor2,thick]
  table[row sep=crcr]{%
0.1	10.3822502302751\\
1	11.6526967029161\\
2	13.3417978716772\\
3	14.4024315755127\\
4	15.1214685021368\\
5	15.6360556528393\\
6	16.0180101885685\\
7	16.3011129452709\\
8	16.4978467791247\\
9	16.6213077821881\\
10	16.7070560494688\\
11	16.7610851383139\\
12	16.7561386519147\\
13	16.753230633592\\
14	16.7345483421303\\
15	16.7193067790635\\
16	16.7000546784558\\
17	16.6939657526492\\
18	16.6960072151423\\
19	16.7073541922483\\
20	16.7183635286325\\
21	16.7344747902759\\
22	16.7455855829866\\
23	16.7521255188458\\
24	16.7641569857502\\
25	16.7564685022901\\
26	16.7463973237639\\
27	16.7387970576132\\
28	16.7472714447747\\
29	16.7421498028213\\
30	16.7344805341179\\
31	16.7443351336617\\
32	16.7460719906698\\
33	16.7459963813536\\
34	16.7563072461376\\
35	16.7625253279766\\
36	16.7586265161012\\
37	16.7578191507131\\
38	16.7536200656986\\
39	16.7489025011633\\
40	16.7490021226143\\
};
\addlegendentry{$R_{\text{opt}}, N_\text{r}^\mathsf{x}=100, \nu D_\text{A,Rx}=\pi$}


\addplot [color=mycolor3, thick]
  table[row sep=crcr]{%
0.1	1.81114474979753\\
1	6.35385539500853\\
2	9.88219398361269\\
3	12.0194626528949\\
4	13.4685316269657\\
5	14.5108325734873\\
6	15.2710330820687\\
7	15.8261322343875\\
8	16.2249189485484\\
9	16.4916813901122\\
10	16.6553985354547\\
11	16.7350090517711\\
12	16.7552962037524\\
13	16.7476148282955\\
14	16.7073808756883\\
15	16.6625732996136\\
16	16.6347772211996\\
17	16.6234871372032\\
18	16.6185525505447\\
19	16.6480518920977\\
20	16.6811094778784\\
21	16.7140013448649\\
22	16.7372405926572\\
23	16.7586055551097\\
24	16.7665160367894\\
25	16.7561125818313\\
26	16.7449908414122\\
27	16.732786608073\\
28	16.7147547992908\\
29	16.7134545783095\\
30	16.7148226095691\\
31	16.7232850781777\\
32	16.7325020074\\
33	16.7394234468531\\
34	16.7404277983572\\
35	16.7512481270531\\
36	16.759769259039\\
37	16.7518807561941\\
38	16.7523648766072\\
39	16.7470766378929\\
40	16.7336624788982\\
};
\addlegendentry{$R_{\text{lin}}, N_\text{r}^\mathsf{x}=100, \nu D_\text{A,Rx}=\pi$}

\addplot [color=mycolor2, dashed, thick]
  table[row sep=crcr]{%
0.1	8.78374439552329\\
1	9.05779096246808\\
2	10.3076422748087\\
3	11.2649236477432\\
4	11.9474014563696\\
5	12.4607444526794\\
6	12.8249761492852\\
7	13.1035221673295\\
8	13.3057977332192\\
9	13.4326593852467\\
10	13.521820816106\\
11	13.5700398001827\\
12	13.5769209770655\\
13	13.5756289007441\\
14	13.5601627708492\\
15	13.5393651381796\\
16	13.5142979630978\\
17	13.5098748488045\\
18	13.5058844414816\\
19	13.5119562163368\\
20	13.5403896974331\\
21	13.5526966718746\\
22	13.562518731743\\
23	13.5827206716264\\
24	13.5811966150381\\
25	13.5727900639278\\
26	13.5809354347917\\
27	13.5671321070419\\
28	13.5617672428072\\
29	13.550594371548\\
30	13.5499764203584\\
31	13.5510959390262\\
32	13.561229109272\\
33	13.5681629202495\\
34	13.5700543016131\\
35	13.5730456542347\\
36	13.5748824169473\\
37	13.5778045276856\\
38	13.5713254118243\\
39	13.5721198969073\\
40	13.5658562323823\\
};
\addlegendentry{$R_{\text{opt}}, N_\text{r}^\mathsf{x}=33, \nu D_\text{A,Rx}=3\pi$}


\addplot [color=mycolor3, thick, dashed]
  table[row sep=crcr]{%
0.1	1.89322612951926\\
1	3.98078243169835\\
2	6.90934996090068\\
3	8.89164052543488\\
4	10.2951471210137\\
5	11.313401656758\\
6	12.0634298985466\\
7	12.6205897703312\\
8	13.0203517445364\\
9	13.2837849136234\\
10	13.4628972466718\\
11	13.5555065227118\\
12	13.5771098427723\\
13	13.5670371629098\\
14	13.5296510733049\\
15	13.4932748654971\\
16	13.4631840893479\\
17	13.4440357138311\\
18	13.4461643175428\\
19	13.4673333216083\\
20	13.4824919905539\\
21	13.5231739320567\\
22	13.5506616257706\\
23	13.5713273735902\\
24	13.5852735002536\\
25	13.5832289706535\\
26	13.5649437328332\\
27	13.5585718125248\\
28	13.5454096788215\\
29	13.5310277114946\\
30	13.533086797455\\
31	13.5421569657452\\
32	13.5447625725515\\
33	13.555650545402\\
34	13.5693566808637\\
35	13.572035776757\\
36	13.5642400888461\\
37	13.5770955621099\\
38	13.5796256311823\\
39	13.5694166109819\\
40	13.5503219625168\\
};
\addlegendentry{$R_{\text{lin}}, N_\text{r}^\mathsf{x}=33, \nu D_\text{A,Rx}=3\pi$}


\end{axis}

\end{tikzpicture}%
		\caption{Fixed \ac{aoa} $\aoa_{\text{mean}}^\text{el}=90^\circ$ and $\NS=2$ satellites}
		\label{fig:RvsDs_th=90}%
	\end{subfigure}
	\hfill
	\begin{subfigure}[b]{0.47\columnwidth}
%
%
\definecolor{mycolor1}{rgb}{0.00000,0.44700,0.74100}%
\definecolor{mycolor2}{rgb}{0.85098,0.32549,0.09804}%
\definecolor{mycolor3}{rgb}{0.46667,0.67451,0.18824}%
\definecolor{mycolor4}{rgb}{0.63529,0.07843,0.18431}%
\definecolor{mycolor5}{rgb}{0.92900,0.69400,0.12500}%
\begin{tikzpicture}

\begin{axis}[%
width=0.85\columnwidth,
height=2in,
scale only axis,
xmin=0,
xmax=100,
xlabel={Inter-satellite  distance $D_\text{S}\,$[km]},
ymin=5,
ymax=28,
axis background/.style={fill=white},
xmajorgrids,
xtick style={color=black},
ymajorgrids,
ytick style={color=black},
legend style={at={(0.98,0.01)}, anchor=south east, legend cell align=left, align=left, fill opacity=0.8, draw opacity=1, text opacity=1, draw=white!80!black}
]
\addplot [color=mycolor5, thick]
  table[row sep=crcr]{%
0.1	1.6942120409075\\
1	2.93552847020509\\
2	3.86721439790162\\
3	4.69630259980829\\
5	6.41206098525234\\
10	12.8640824021329\\
15	17.9377483643994\\
20	20.4331765785548\\
25	22.2427682621381\\
30	23.5010798719901\\
35	24.4508008404084\\
40	25.1855566091792\\
45	25.7296361915748\\
50	26.1414820654196\\
55	26.3917179957027\\
60	26.5940549304578\\
65	26.6908837504284\\
70	26.7266589681665\\
75	26.7406707566294\\
80	26.7285197713445\\
85	26.7346360867586\\
90	26.7494558200946\\
95	26.7683263249623\\
100	26.7310485085255\\
};
\addlegendentry{$N_\text{S}=6$}

\addplot [color=mycolor4, thick]
  table[row sep=crcr]{%
0.1	1.72077996776147\\
1	2.96890576527628\\
2	3.84734661290048\\
3	4.69024942016101\\
5	6.37959676079481\\
10	12.7114798061615\\
15	17.0435139293743\\
20	19.2199926273798\\
25	20.8647198541824\\
30	21.9822877437796\\
35	22.8614656097213\\
40	23.4896193497616\\
45	24.0025815556951\\
50	24.3290342749511\\
55	24.6198579310883\\
60	24.7374051490195\\
65	24.8006655325068\\
70	24.8197485934546\\
75	24.837362829871\\
80	24.8607542216409\\
85	24.8477424741313\\
90	24.8295357831085\\
95	24.8244315091019\\
100	24.8454540872736\\
};
\addlegendentry{$N_\text{S}=5$}

\addplot [color=mycolor3, thick]
  table[row sep=crcr]{%
0.1	1.7685266148093\\
1	3.05067730379879\\
2	3.84100848037288\\
3	4.76504548099596\\
5	6.40519517899094\\
10	12.4991614400864\\
15	15.9915764869897\\
20	17.8600173850997\\
25	19.2194344471046\\
30	20.1618780637682\\
35	20.8747555295356\\
40	21.4602310466967\\
45	21.8581985210881\\
50	22.1354936034566\\
55	22.2599353048694\\
60	22.350172058509\\
65	22.357929498768\\
70	22.3780793325142\\
75	22.3875420614068\\
80	22.3808428763157\\
85	22.3737360827156\\
90	22.3833975834698\\
95	22.3889416807403\\
100	22.3945914094739\\
};
\addlegendentry{$N_\text{S}=4$}

\addplot [color=mycolor2, thick]
  table[row sep=crcr]{%
0.1	1.85619601141289\\
1	3.2135296793967\\
2	4.04627228876016\\
3	4.90030746220218\\
5	7.2572591573855\\
10	12.3046190855951\\
15	14.8090355429763\\
20	16.2399730916854\\
25	17.2968195807193\\
30	17.9948665388374\\
35	18.4943956430293\\
40	18.7955146572612\\
45	18.9972897364198\\
50	19.1361537909865\\
55	19.2065000336494\\
60	19.238289328316\\
65	19.2482802419706\\
70	19.257294542637\\
75	19.2463389776386\\
80	19.2501505490286\\
85	19.2563645180998\\
90	19.253431986025\\
95	19.248051070901\\
100	19.2489806694663\\
};
\addlegendentry{$N_\text{S}=3$}

\addplot [color=mycolor1, thick]
  table[row sep=crcr]{%
0.1	2.05242791194976\\
1	3.5438683868294\\
2	5.56047615692176\\
3	7.1289449919826\\
5	9.35351383829474\\
10	12.3061657368595\\
15	13.5523398894074\\
20	14.1807736816677\\
25	14.5786222056965\\
30	14.7918834639706\\
35	14.9553094726255\\
40	15.0380579826921\\
45	15.1021854568484\\
50	15.1364934058906\\
55	15.1501642303162\\
60	15.1645418101488\\
65	15.1751367405838\\
70	15.1693822116931\\
75	15.1661459995904\\
80	15.1615863272002\\
85	15.1879044958225\\
90	15.1613264241738\\
95	15.1724278019557\\
100	15.1670659230736\\
};
\addlegendentry{$N_\text{S}=2$}

\addplot [color=black, thick]
  table[row sep=crcr]{%
0.1	9.57291182570065\\
100	9.56673367166231\\
};
\addlegendentry{$N_\text{S}=1$}

\end{axis}

\end{tikzpicture}%
		\caption{Averaged over time} 
		\label{fig:RvsDs_avrg}%
	\end{subfigure}
	\caption{Achievable rate performance in dependence of different inter-satellite distances $\DS$ with $\aoal^{\text{az}}=0$ for all $\ell$.} 
\end{figure}
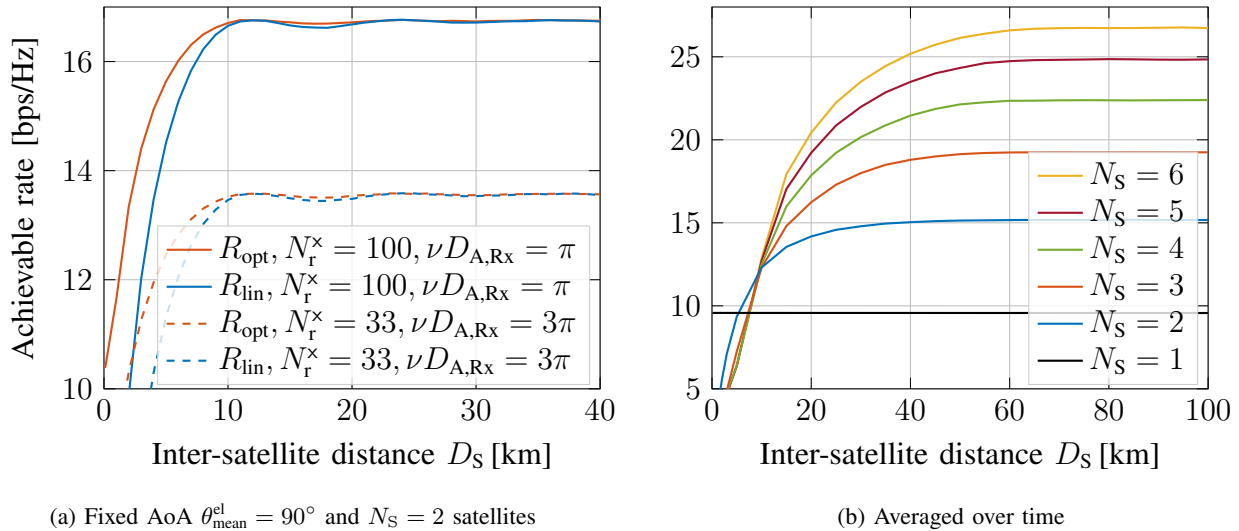

In Fig. \ref{fig:RvsDs_th=90}, the achievable rate is shown as a function of the inter-satellite distance $\DS$ with a fixed average elevation angle $\aoa_{\text{mean}}^{\text{el}}=90^\circ$.
Furthermore, two different antenna arrays at the \ac{rx} have been assumed in Fig. \ref{fig:RvsDs_th=90}. The solid lines are for an array with $\Nr^\mathsf{x}=100$ receive antennas and an antenna spacing of half of the wavelength, i.e., $\nu\DARx=\pi$, which is also assumed in the rest of this paper. For the dashed lines, the number of receive antennas is $\Nr^\mathsf{x}=33$, while the antenna spacing is $1.5$ times the carrier wavelength, i.e., $\nu\DARx=3\pi$.
According to the derivation from Section \ref{sec:d_sat}, the achievable rate must have its maximum every $D_{\text{S,opt}}=\SI{12}{\kilo\meter}$, for both receive apertures.
It can  be seen, that this is fulfilled \rew{for the channel capacity $R_\text{opt}$ as well as the achievable rate with our proposed linear transceiver approach $R_\text{lin}$.
For inter-satellite distances $\DS>\SI{12}{\kilo\meter}$, the achievable rate}
oscillates slightly, while every $D_{\text{S,opt}}=\SI{12}{\kilo\meter}$ the maximum is obtained. %
The overall rate for $\Nr^\mathsf{x}=33$ is less than with $\Nr^\mathsf{x} = 100$ antennas, which is due to the smaller array gain.
Considering the proposed precoding and equalization approach, one can observe, that 
the capacity is achieved at the optimal inter-satellite distances.
Additionally, for $\DS>\SI{12}{\kilo\meter}$, the achievable rate is still close to the optimum.
However, for $\DS<\SI{12}{\kilo\meter}$, the proposed approach leads to a rate reduction compared to optimal transceivers. This is mainly due to the linear equalization and could be alleviated by using \ac{sic} at the \ac{rx}.

In real world applications, the time averaged rate is of greater interest than the rate at a fixed time instance.
Therefore, in Fig. \ref{fig:RvsDs_avrg}, the achievable rate is averaged over a complete satellites pass. Then, there are no more periodic variations of the achievable rate, but the achievable rate stays constant for $\DS\ge \SI{65}{\kilo\meter}$.
This corresponds to the optimum inter-satellite distance at the minimum elevation angle $\aoa_{\text{mean,min}}^{\text{el}}=30^\circ$ and holds for all considered number of satellites $\NS\in\{2,3,4,5,6\}$, which justifies the heuristic \eqref{eq:orth_cond_relaxed}.

\subsection{Imperfect Position Knowledge}\label{sec:sim_robust}
\begin{figure}[t]%
	\centering
	\begin{subfigure}{0.47\columnwidth}
%
%
\definecolor{mycolor0}{rgb}{0.85000,0.32500,0.09800}%
\definecolor{mycolor1}{rgb}{0.00000,0.44700,0.74100}%
\definecolor{mycolor2}{rgb}{0.49400,0.18400,0.55600}%
\definecolor{mycolor3}{rgb}{0.46667,0.67451,0.18824}%
\begin{tikzpicture}

\begin{axis}[%
width=0.85\columnwidth,
height=2in,
scale only axis,
xmin=0,
xmax=30,
xlabel style={font=\color{white!15!black}},
xlabel={Sum transmit power $P_{\text{Tx}}\,$[dBW]},
ymin=0,
ymax=35,
ylabel style={font=\color{white!15!black}},
ylabel={Achievable Rate [bps/Hz]},
axis background/.style={fill=white},
xmajorgrids,
ymajorgrids,
legend style={at={(0.01,0.98)}, anchor=north west, legend cell align=left, align=left, fill opacity=0.8, draw opacity=1, text opacity=1, draw=white!80!black}
]
\addplot [color=mycolor0, thick]
  table[row sep=crcr]{%
0	10.7438170364659\\
5	13.9990297267307\\
10	17.29935463194\\
15	20.6143979695948\\
20	23.9341435167496\\
25	27.2553808814992\\
30	30.5770904970767\\
};
\addlegendentry{$\text{R}_{\text{opt}}$}

\addplot [color=mycolor1, dashed, thick, mark=+, mark options={solid}]
  table[row sep=crcr]{%
0	10.7242783960862\\
5	13.9763623632696\\
10	17.2754571058872\\
15	20.5900835797706\\
20	23.9096933085791\\
25	27.2308839523385\\
30	30.5525677456254\\
};
\addlegendentry{$\upsilon_{\text{max}}=0, \xi_{\text{max}}=0$}

\addplot [color=mycolor2, thick, mark=o, mark options={solid, mycolor2}]
  table[row sep=crcr]{%
0	3.11445583651938\\
5	4.65036086669475\\
10	6.35816799224874\\
15	8.31704029644432\\
20	10.1155553378069\\
25	11.9615187921995\\
30	13.7009115898947\\
};
\addlegendentry{Robust, $\upsilon_{\text{max}}= \frac{1}{64}$}

\addplot [color=mycolor3, thick, dashed, mark=o, mark options={solid, mycolor3}]
  table[row sep=crcr]{%
0	1.52197578146014\\
5	2.52110322800324\\
10	3.65240826301087\\
15	4.97592498103996\\
20	6.2628009404809\\
25	7.61397854146529\\
30	8.66345005928268\\
};
\addlegendentry{Heuristic, $\upsilon_{\text{max}}= \frac{1}{64}$}

\addplot [color=mycolor2, thick, mark=triangle, mark options={solid, rotate=180, mycolor2}]
  table[row sep=crcr]{%
0	0.666323121083333\\
5	1.06785227809254\\
10	1.55076820256642\\
15	2.1605536872615\\
20	2.79752838167133\\
25	3.45875573404426\\
30	4.15490104667279\\
};
\addlegendentry{Robust, $\upsilon_{\text{max}}= \frac{1}{32}$}

\addplot [color=mycolor3, dashed, thick, mark=triangle, mark options={solid, rotate=180, mycolor3}]
  table[row sep=crcr]{%
0	0.513438738274257\\
5	0.937541716448588\\
10	1.4899328564898\\
15	2.19606695206833\\
20	3.13888154500095\\
25	4.00685565290786\\
30	4.91807071967802\\
};
\addlegendentry{Heuristic, $\upsilon_{\text{max}}= \frac{1}{32}$}

\end{axis}
\end{tikzpicture}%
		\caption{\rew{$\DS=\SI{26}{\kilo\meter}$}}
		\label{fig:robust_d_small}
	\end{subfigure}
	\hfill
	\begin{subfigure}{0.47\columnwidth}
%
%
\definecolor{mycolor0}{rgb}{0.85000,0.32500,0.09800}%
\definecolor{mycolor1}{rgb}{0.00000,0.44700,0.74100}%
\definecolor{mycolor2}{rgb}{0.92941,0.69412,0.12549}%
\definecolor{mycolor3}{rgb}{0.49400,0.18400,0.55600}%
\definecolor{mycolor4}{rgb}{0.46667,0.67451,0.18824}%
\begin{tikzpicture}

\begin{axis}[%
width=0.85\columnwidth,
height=2in,
scale only axis,
xmin=0,
xmax=30,
xlabel style={font=\color{white!15!black}},
xlabel={Sum transmit power $P_{\text{Tx}}\,$[dBW]},
ymin=0,
ymax=35,
ylabel style={font=\color{white!15!black}},
ylabel={Achievable Rate [bps/Hz]},
axis background/.style={fill=white},
xmajorgrids,
ymajorgrids,
legend style={at={(0.01,0.98)}, anchor=north west, legend cell align=left, align=left, fill opacity=0.8, draw opacity=1, text opacity=1, draw=white!80!black}
]
\addplot [color=mycolor0, thick]
  table[row sep=crcr]{%
0	10.898062766981\\
5	14.1610035228688\\
10	17.4639302059541\\
15	20.7798134010712\\
20	24.0998262763828\\
25	27.421148364429\\
30	30.7428847846254\\
};
\addlegendentry{$\text{R}_{\text{opt}}$}

\addplot [color=mycolor1, dashed, thick, mark=+, mark options={solid}]
  table[row sep=crcr]{%
0	10.8937666629988\\
5	14.1567911422722\\
10	17.4596788976372\\
15	20.7755423740787\\
20	24.0955480605787\\
25	27.4168672049425\\
30	30.7386008320471\\
};
\addlegendentry{$\upsilon_{\text{max}}=0, \xi_{\text{max}}=0$}

\addplot [color=mycolor3, thick, mark=o, mark options={solid}]
  table[row sep=crcr]{%
0	3.58208122673572\\
5	5.58852724979415\\
10	8.00233285849046\\
15	10.5943611789471\\
20	13.241602079746\\
25	16.1910673711038\\
30	18.8704884903681\\
};
\addlegendentry{Robust, $\upsilon_{\text{max}}= \frac{1}{64}$}

\addplot [color=mycolor4, dashed, thick, mark=o, mark options={solid}]
  table[row sep=crcr]{%
0	1.55207703875706\\
5	2.62713200277245\\
10	3.98700438735168\\
15	5.60898624111355\\
20	7.27455458314958\\
25	9.20923643780615\\
30	10.8398450964287\\
};
\addlegendentry{Heuristic, $\upsilon_{\text{max}}= \frac{1}{64}$}

\addplot [color=mycolor3, thick, mark=triangle, mark options={solid, rotate=180}]
  table[row sep=crcr]{%
0	1.26220180480334\\
5	2.10817900847533\\
10	3.20496427963359\\
15	4.54322663400543\\
20	6.20692015710071\\
25	7.73383382711054\\
30	9.69790000831241\\
};
\addlegendentry{Robust, $\upsilon_{\text{max}}= \frac{1}{32}$}

\addplot [color=mycolor4, dashed, thick, mark=triangle, mark options={solid, rotate=180}]
  table[row sep=crcr]{%
0	0.506426216634168\\
5	0.979097358324126\\
10	1.6049667987731\\
15	2.44296194005055\\
20	3.53251448443409\\
25	4.80664889587023\\
30	6.06927440202298\\
};
\addlegendentry{Heuristic, $\upsilon_{\text{max}}= \frac{1}{32}$}

\end{axis}

\end{tikzpicture}%
		\caption{\rew{$\DS=\SI{52}{\kilo\meter}$}}
		\label{fig:robust_d_opt}
	\end{subfigure}
	\caption{\rew{Achievable rate performance for robust precoder and equalizer.}}
	\label{fig:robust_txrx}%
\end{figure}
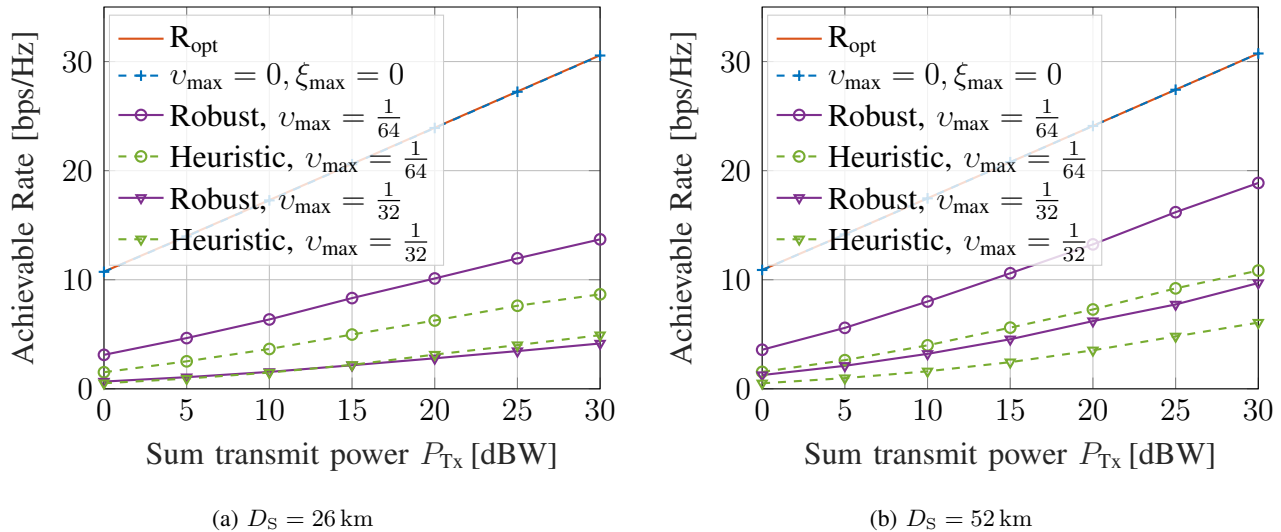

\rew{Next, the performance of the proposed precoder and equalizer is evaluated under more practical assumptions. We assume $\NS=2$ satellites with a constant inter-satellite distance $\DS\in\{\SI{26}{\kilo\meter}, \SI{52}{\kilo\meter}\}$, each equipped with an $8\times 8$ \ac{ura}, while the \ac{rx} has an $16\times 16$ \ac{ura}. The corresponding antenna gains are $\GTx=\SI{17.4}{\dBi}$ and $\GRx=\SI{15.6}{\dBi}$, respectively, and the antenna spacing for all arrays is $\DATx=\DARx=\SI{6}{\centi\meter}$. In this case, the overall array gains, i.e., $\GTx+10\log_{10}(\NS\Nt)=\SI{38.5}{\dBi}$ and $\GRx+10\log_{10}(\Nr)=\SI{39.7}{\dBi}$, matches with the 3GPP recommendations \cite{3GPP.TR.38.821}.
The corresponding achievable rates are shown in Fig.~\ref{fig:robust_txrx} w.r.t. the sum transmit power $\Ptx$.
In both graphs, uniform estimation errors of the space angles at the satellites and the \ac{rx} are assumed, i.e., $\erraod_\ell^\mathsf{x},\erraod_\ell^\mathsf{y}\sim \mathcal{U}(-\erraod_{\max}, \erraod_{\max})$ and $\erraoa_\ell^\mathsf{x},\erraoa_\ell^\mathsf{y}\sim \mathcal{U}(-\erraoa_{\max}, \erraoa_{\max})$, with $\erraod_{\max}=1/16$ and $\erraoa_{\max}\in\{1/16, 1/32\}$.
The purple lines show the performance of the proposed robust transceiver from Section \ref{sec:impCSI}. The green lines show the performance of a heuristic approach, where the precoder and equalizer are calculated with \eqref{eq:pc_ad} and \eqref{eq:eq_geo}, respectively, by substituting the true angles with the estimated ones. As benchmark, the channel capacity $R_\text{opt}$ is included. 
It can be seen that by including the statistic of the error, the performance degradation due to imperfect position knowledge is reduced.
According to previous observations, the optimal inter-satellite for $\Nr^\mathsf{x}=16$ and $\nu\DARx=8\pi$ should be at $D_{\text{S,opt}}\approx\SI{52}{\kilo\meter}$. Given that inter-satellite distance, the channel capacity is achieved with the proposed transceiver approach, under perfect position knowledge, i.e., $R_\text{opt}\approx R_\text{lin}$, for both the heuristic and the robust one. 
Thus, given the proposed linear precoder and equalizer, the channel capacity can  be achieved by transmitting independent data streams, which avoids time-critical inter-satellite communication, and without the necessity of \ac{sic} at the \ac{rx}.
Further increasing the inter-satellite distance does not increase the capacity $R_\text{opt}$ nor the achievable rate $R_\text{lin}$. However, as observed in Fig. \ref{fig:robust_d_small}, even at lower inter-satellite distances the same rate can be achieved. In case of imperfect position knowledge, the performance with a smaller inter-satellite distance decreases stronger than for larger distances. }

\section{Conclusion}\label{sec:conclusion}
In this paper, a low complexity distributed precoder for satellite swarms and a linear equalizer, both utilizing the geometric relation between the positions of the satellites and \ac{rx}, has been proposed.
\rew{Furthermore, the inter-satellite distance has been optimized for a specific scenario.}
Given that the inter-satellite distances are chosen adequately, it has been shown that the proposed transceiver architecture achieves a performance very close to the capacity upper bound obtained by assuming perfect CSI and instantaneous coordination between satellites. \rew{This almost optimal performance is achieved without any inter-satellite communication between the satellites, for the precoder design.}
Furthermore, the requirements for channel estimation are significantly reduced, as only relative positional knowledge between the satellites and the \ac{rx} are necessary. 
If the relative positions are estimated erroneously, knowledge about the statistic of the estimation error can be exploited to reduce the performance degradation.
However, not all issues in real world scenarios have been considered in this paper, e.g.,  finite length channel codes and discrete modulation schemes must be used in practical communications \rew{and the \ac{los} connection can be blocked}. This effects will lead to an overall performance degradation in over-the-air transmission.
Nevertheless, it has been shown that in the downlink, by applying the proposed approaches, the achievable rate can be significantly increased for satellite swarms compared to single satellite systems, while keeping the computational complexity and \rew{signaling overhead} for precoding and equalization low.

\appendices
\section{Generalized Rayleigh Quotient} \label{sec:rayleigh}
\begin{lemma} \label{lem:rayleigh}
Let $\bm{A}\in \mathbb{C}^{n\times n}$ be Hermitian and $\bm B\in \mathbb{C}^{n\times n}$ Hermitian positive definite. Then,
\begin{equation} \label{eq:rayleigh}
    \max_{\bm{x}\in\mathbb C^n\setminus\{\bm{0}\}} \frac{\bm x^H \bm{A x}}{\bm x^H \bm{B x}} = \lambda_\mathrm{max}(\bm A, \bm B),
\end{equation}
where $\lambda_\mathrm{max}(\bm A, \bm B)$ is the maximum generalized eigenvalue. The optimal $\bm x$ in \eqref{eq:rayleigh} is an eigenvector corresponding to $\lambda_\mathrm{max}(\bm A, \bm B)$.
\end{lemma}

\begin{IEEEproof}
Let $\bm{LL}^H$ be the Cholesky factorization of $\bm B$. Substitute $\bm y = \bm L^H \bm x$ in the LHS of \eqref{eq:rayleigh}. Then,
\begin{equation}
    \max_{\bm{x}\in\mathbb C^n\setminus\{\bm{0}\}} \frac{\bm x^H \bm{A x}}{\bm x^H \bm{LL}^H \bm x}
    = \max_{\bm{x}\in\mathbb C^n\setminus\{\bm{0}\}}  \frac{\bm y^H \bm L^{-1} \bm {A L}^{-H} \bm y}{\bm y^H \bm y}
    = \lambda_\mathrm{max}(\bm L^{-1} \bm {A L}^{-H}),
\end{equation}
where the last step is due to the Rayleigh-Ritz Theorem \cite[Thm.~4.2.2]{Horn1990}.

Recall that two square matrices $\bm C, \bm D$ are similar if there exists a nonsingular matrix $\bm S$ such that $\bm D = \bm S^{-1} \bm{C S}$. By virtue of \cite[Cor.~1.3.4]{Horn1990}, similar matrices have the same eigenvalues. Since $\bm L^{-1} \bm {A L}^{-H}$ and $ (\bm{L L}^{H})^{-1} \bm A$ are similar with similarity matrix $\bm L^{H}$, $\lambda_\mathrm{max}(\bm L^{-1} \bm {A L}^{-H}) = \lambda_\mathrm{max}(\bm B^{-1} \bm A) = \lambda_\mathrm{max}(\bm A, \bm B)$.

Let $\tilde{\bm y}$ be an eigenvector of $\bm L^{-1} \bm {A L}^{-H}$ corresponding to the maximum eigenvalue $\lambda_\mathrm{max}$. Then, by \cite[Def.~1.1.2]{Horn1990}, $\bm L^{-1} \bm {A L}^{-H} \tilde{\bm y} = \lambda_\mathrm{max} \tilde{\bm y}$. Multiplying both sides from the left by $\tilde{\bm y}^H$ and substituting $\tilde{\bm y} = \bm L^H \tilde{\bm x}$, we obtain
\begin{align}
    (\bm L^H \tilde{\bm x})^H \bm L^{-1} \bm {A L}^{-H} (\bm L^H \tilde{\bm x}) = \lambda_\mathrm{max} (\bm L^H \tilde{\bm x})^H (\bm L^H \tilde{\bm x}) \quad\Leftrightarrow\quad
    \tilde{\bm x}^H \bm A \tilde{\bm x} = \lambda_\mathrm{max} \tilde{\bm x}^H \bm B \tilde{\bm x}.
\end{align}
This establishes that $\bm L^H \tilde{\bm x}$ maximizes $\bm x^H \bm{A x} / (\bm x^H \bm{B x})$. Due to similarity and the fact that $\tilde{\bm y}$ is an eigenvector of $\bm L^{-1} \bm {A L}^{-H}$ corresponding to $\lambda_\mathrm{max}$, $\tilde{\bm x}$ is an eigenvector of $\bm B^{-1} \bm A$ \cite[Thm.~1.4.8]{Horn1990}.
\end{IEEEproof}

\section{Explicit formula for $\D_{\text{S,opt}}$}\label{sec:D_S}
For the sake of brevity, only the case $\minrew{\cos(\aoal^\mathrm{el})\ge \cos(\aoa_{\ell-1}^\mathrm{el})}$ is considered here. The formula for the other case can be found in the same way.
With \eqref{eq:theta_l-1} and \eqref{eq:vartheta}, $\minrew{\vartheta_{\ell-1}^\mathrm{el}}$ can be written as a function of the \minrew{elevation angle  $\aoal^\mathrm{el}$} of satellite $\ell$ and the receive array parameters \minrew{$\nu\DARx\Nr^\mathsf{x}$}
\minrew{\begin{align}
   \vartheta_{\ell-1}^\mathrm{el} = \arccos\left(\cos(\aoal^\mathrm{el}) - \frac{2\pi}{\nu\DARx\Nr^\mathsf{x}}\right) +
   \arcsin\left(\frac{\RE}{r_0}\left(\cos(\aoal^\mathrm{el}) - \frac{2\pi}{\nu\DARx\Nr^\mathsf{x}}\right) \right)\,.
\end{align}}
Now, with \eqref{eq:d_sin}, the distances $d_\ell$ and $d_{\ell-1}$ can be written as a function of $\minrew{\aoal^\mathrm{el}}$, the receive array parameters $\minrew{\nu\DARx\Nr}$ as well as the orbital radius $r_0$
\minrew{\begin{subequations}\label{eq:d_l_long}
 \begin{align}
    d_{\ell-1} &= r_0\frac{\cos\left( \arccos\left(\cos(\aoal^\mathrm{el}) - \frac{2\pi}{\nu\DARx\Nr^\mathsf{x}}\right) +
   \arcsin\left(\frac{\RE}{r_0}\left(\cos(\aoal^\mathrm{el}) - \frac{2\pi}{\nu\DARx\Nr^\mathsf{x}}\right)\right) \right)}{\cos\left(\aoal^\mathrm{el}\right) - \frac{2\pi}{\nu\DARx\Nr^\mathsf{x}}}
    \\
    d_{\ell} &= r_0\frac{\cos\left(\aoal^\mathrm{el} + \arcsin\left(\frac{\RE}{r_0} \cos(\aoal^\mathrm{el})\right)\right)}{\cos(\aoal^\mathrm{el})}\,.
\end{align}   
\end{subequations}}
Finally, plugging \eqref{eq:d_l_long} into \eqref{eq:D_S} gives the smallest optimal inter-satellite distance $D_\text{S,opt}$
\minrew{
{\small\begin{flalign}\label{eq:D_S_opt}
        D_\text{S,opt} =  r_0^2
        \bigg[ 
        &\bigg(\frac{\cos\left(\aoal^\mathrm{el} + \arcsin\left(\frac{\RE}{r_0} \cos(\aoal^\mathrm{el})\right)\right)}{\cos(\aoal^\mathrm{el})}\bigg)^2 \notag\\
        & +\bigg(
        \frac{\cos\left( \arccos\left(\cos(\aoal^\mathrm{el}) - \frac{2\pi}{\nu\DARx\Nr^\mathsf{x}}\right) +
        \arcsin\left(\frac{\RE}{r_0}\left(\cos(\aoal^\mathrm{el}) - \frac{2\pi}{\nu\DARx\Nr^\mathsf{x}}\right) \right) \right)}{\cos\left(\aoal^\mathrm{el}\right) - \frac{2\pi}{\nu\DARx\Nr^\mathsf{x}}}\bigg)^2 \notag\\
   & - 2 \frac{\cos\left(\aoal^\mathrm{el} + \arcsin\left(\frac{\RE}{r_0} \cos(\aoal^\mathrm{el})\right)\right)}{\cos(\aoal^\mathrm{el})} \notag\\
   &\phantom{{}-{}}\cdot 
    \frac{\cos\left( \arccos\left(\cos(\aoal^\mathrm{el}) - \frac{2\pi}{\nu\DARx\Nr^\mathsf{x}}\right) +
   \arcsin\left(\frac{\RE}{r_0}\left(\cos(\aoal^\mathrm{el}) - \frac{2\pi}{\nu\DARx\Nr^\mathsf{x}}\right) \right) \right)}{\cos\left(\aoal^\mathrm{el}\right) - \frac{2\pi}{\nu\DARx\Nr^\mathsf{x}}} \notag\\
   &&\mathllap{\cdot \cos\left(\arccos\left(\cos\left(\aoal^\mathrm{el}\right) - \frac{2\pi}{\nu\DARx\Nr^\mathsf{x}}\right) - \aoal^\mathrm{el} \right)}
   \bigg]^{\frac{1}{2}}\!\!.
\end{flalign}}
}

%

\bibliographystyle{./lib/IEEEtran}
\bibliography{./lib/IEEEabrv,./lib/references_short}

\end{document}